\documentclass[%
reprint,
 amsmath,amssymb,
 aps,
pra,
]{revtex4-2}

\usepackage{CJKutf8}
\usepackage{graphicx}
\usepackage{dcolumn}
\usepackage{bm}
\usepackage{hyperref}

\usepackage{xcolor}
\newcommand{\Y}[1]{{\color{blue} {\bf} #1}}

\begin{document}

\begin{CJK}{UTF8}{gbsn}

 \title{Breakdown effect of periodic perturbations to the robustness of topological phase\\ in a gyromagnetic photonic crystal}

\author{{Y. Tian(田玉)}\textsuperscript{1, 2}}
\author{{Rui Zhou}\textsuperscript{3}}
\author{{Zheng-Rong Liu(刘峥嵘)}\textsuperscript{1}}
 \author{{Y. Liu(刘泱杰)}\textsuperscript{1, 2, 4}}
\email{Corresponding author: yangjie@hubu.edu.cn}

\author{{Hai Lin}\textsuperscript{3}}
\email{Corresponding author: linhai@mail.ccnu.edu.cn}
\author{{Bin Zhou}\textsuperscript{1}}

\affiliation{
$^{1}$ School of Physics, Hubei University, Wuhan 430062, Hubei Province\\
$^{2}$ School of Micro-Electronics, Hubei University, Wuhan 430062, Hubei Province\\
$^{3}$ College of Physics Science and Technology, Central China Normal University, Wuhan 430079, Hubei Province\\
$^{4}$ Lanzhou Center for Theoretical Physics, Key Laboratory of Theoretical Physics of Gansu Province, and Key Laboratory of Quantum Theory and Applications of MoE, Lanzhou University, Lanzhou 730000, Gansu Province
}%

\date{\today}

\begin{abstract}
~\footnote{Submitted to \itshape{Phys. Lett. A} on 29th Mar.'23, PHYLA-D-23-00377R, reviewer reports received 19th Apr., revised 4th, 18th May, resubmitted 29th Mar., rejected 12th Jun. , transferred to  \itshape{Results in Physics} on 4 Jul., RINP-D-23-02097 and rejected 18th Jul. 2023. Submitted to IET Optoelectronics 26th July, major revision decided 5th Sept., revision submitted 17th, and accepted 24th Sept. as OPT-2023-07-0033.R1. }
In the known field of topological photonics, what remains less so is the breakdown effect of topological phases deteriorated by perturbation. In this paper, we investigate the variance on topological invariants for a periodic Kekul{\'e} medium perturbed in unit cells, which was a gyromagnetic photonic crystal holding topological phases induced by \emph{synchronized rotation} of unit cells. Two parameters for geometric and material perturbation are respectively benchmarked to characterise the topological degradation. Our calculation demonstrates that such a periodic perturbation easily destructs the  topological phase, and thus calls for further checkups on robustness under such unit-cell-perturbation in realization. 
\end{abstract}

\maketitle
\end{CJK}

\section{\label{sec1}Introduction}

The mathematical concept of topology, referring to which attribute of geometrical objects remains unchanged after continuous transformation, has inspired the learned society in physics to investigate both quantum systems and classical analogues to identify their novel topological phenomenon~\cite{Kane2005a, Haldane2008, WangZ2008, Hasan2010, Khanikaev2012, WuL2015, Ozawa2019, Price2022}. One of the promisingly applicable feature of such systems relies on the non-zero topological invariants defined in reciprocal space in a periodic medium, of which the unit cells are judiciously designed to occupy certain symmetry, in order to realize scattering-free directional flows along the interface between distinct matters~\cite{WangZ2009, Fu2011, LiangG2013, Rechtsman2013, WuL2015, XuL2016, WenX2018, GengY2019, HeC2020, XieL2023}. This unprecedented scattering-less feature is just saliently dictated by the bulk-edge correspondence as a result of such topological insights. 

In emergence of these efforts to contribute to practical design of topological photonic crystals (PhC), \emph{synchronized rotation} was proposed as a design trick to induce topological phases, namely rotating the whole unit cell synchronously all over the periodic lattice~\cite{ZhouR2021, WangX2020, ChenJ2022, Joannopoulos2008, SuiY2022}. This trick essentially adjusts the intercell and intracell coupling strengths resulting to implicit tunable phases, which generalizes from the simple picture of Su–Schrieffer–Heeger (SSH) chain~\cite{ZhouR2021, ZhouR2022, YuZ2022}. In the hope to build for an amenable photonic platform towards future semiconductor industry, pure dielectrics rather than conductors is called in, to provide a reconfigurable and implementable testbed for the fields of topological quantum physic~\cite{WuL2015, YangY2018}. One of the core metrics to measure topological characteristics is Chern number as a topological invariant integer in definition~\cite{Hasan2010, WuL2015}.

If such a periodic lattice distorts under effect of geometric randomness, disorder shall come into play and greatly complicate the topological picture. First, a local disorder does not change the global topological phase, against which topologically-protected edge states remain robust. Second, when disorder exerts global randomness, it destroys the periodic regularity and the topologically-protected edge states ceases into scattering. And as the disorder pushes further, it may surprisingly induce phase transitions via onsite disorder, which was named as \emph{topological Anderson effect}~\cite{LiJ2009, LiuC2017, Stutzer2018, Meier2018}. For a disordered medium globally-randomized away from its original periodic one, its topological invariant can be defined as Bott index~\cite{LiuG2020}. Nevertheless, for non-periodic continuous medium, this could be defined otherwise practically or theoretically, \emph{e.g.} in various metamaterial media~\cite{LiuC2017, CuiX2022, Silveirinha2015}. 

Furthermore, a periodic lattice can also occupy another type of disorder which may break the originally-held spatial symmetries, by disrupting each unit cell ~\emph{uniformly} across the whole lattice. This type of disorder occurs only within the unit cell while maintaining periodicity for ease of calculation. The consequence of this type of perturbation to topological transition should be vital in consideration of pragmatic manufacturing of topological PhC and thus worth our efforts to pinpoint the breakdown effect of topological protection. Here we investigate whether such effects occur in the context of gyromagnetic PhC~\cite{WangZ2008, ChenZ2017, ChenJ2020, RaoS2022, SuiY2023} where randomness is brought uniformly within only the unit cells. In this paper, we calculate the Chern numbers in our classical wave system to mark up how the topological phases are influenced by \emph{periodic perturbation} concretely. Moreover, we move forward to question the robustness of the protected edge states, and to pinpoint the topological Anderson effect when such-made PhC structure is under geometric randomness in unit cells. We choose two configurable parameters for geometric and material perturbation, to characterize the topological breakdown via directly calculating Chern numbers under our perturbation, which are induced in a most natural manner without making complicated mapping. Finally, we find that the topological invariant is easily annihilated by our periodic perturbation, providing benchmark for future design of these topological devices.

\section{\label{sec2} Regular lattice before disruption: our model and numeric calculation of Chern numbers}

\begin{figure}
\includegraphics[width=0.45\textwidth]{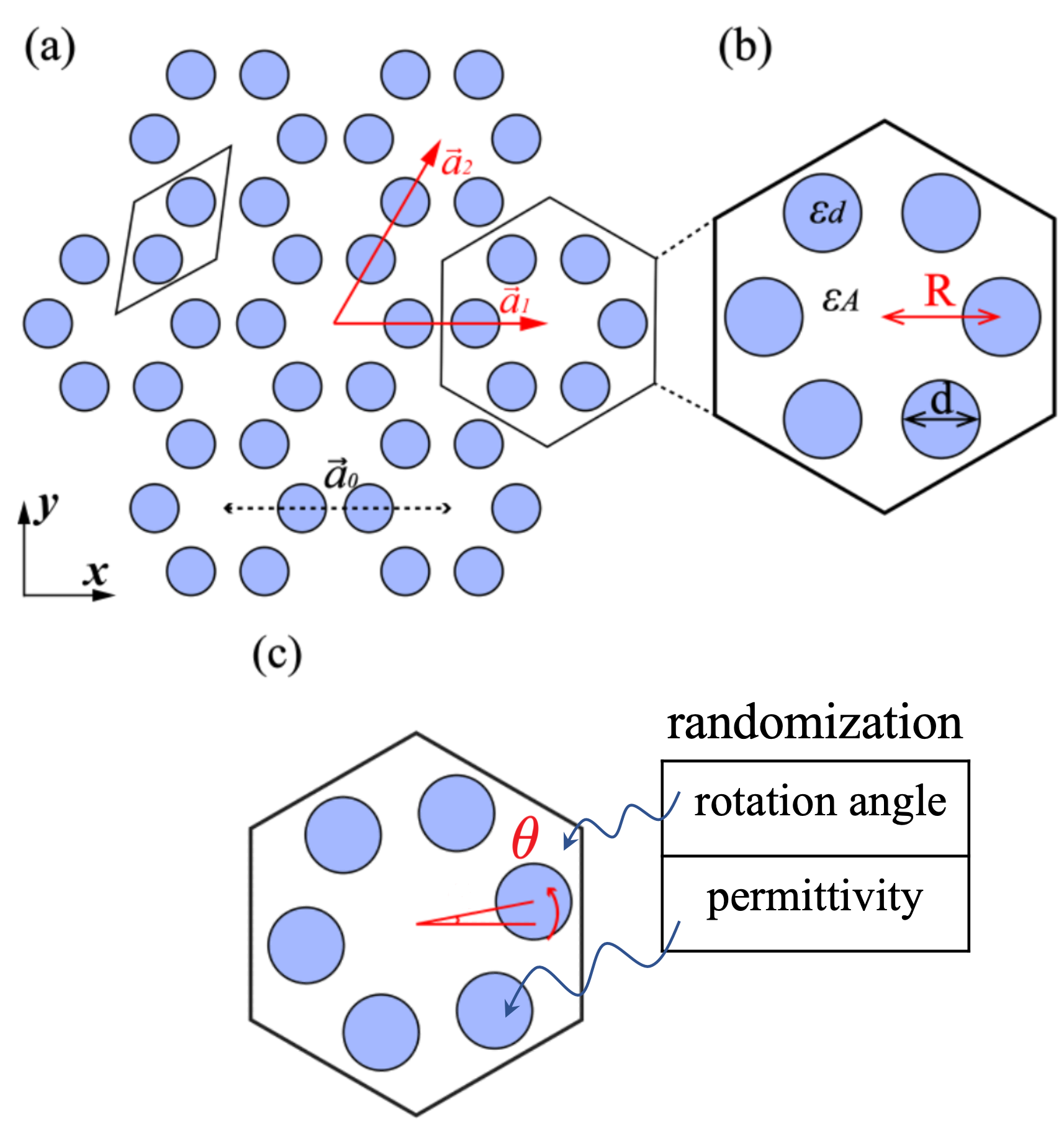}
\caption{\label{fig1} Schematic diagram of a two-dimensional Kekul\'e gyromagnetic topological PhC with a lattice constant of ${a}_{0}$. (a) The arrangement of the lattice unit marked by the black solid line and the lattice red vectors. (b) Unit cell before rotation with $R$ as the distance from the centre of unit cell to that of each circular pillar. (c) Unit cell under a rotation angle $\theta=9^{\circ}$ for instance. In this paper, periodic perturbation is induced by randomizing two structural parameters: rotation angle and pillar permittivity in unit cells. } 
\end{figure}

In this work, we use a finite-element numeric method to calculate the PhC dispersion and also the electric field solution for our Kekul\'e gyromagnetic photonic crystals whose topological phase is induced by synchronized rotation. Indicated by Fig.~\ref{fig1}(a-b), each unit cell of a hexagon is displaced by a lattice constant $a_0=1$, and made of six cylindrical pillars of diameter $d=0.11a_0, R=a_0/2.92, \varepsilon_{\rm d}=15.0$ throughout our paper, where $R$ is the distance from the centre of unit cell to that of each circular pillar and $a$ the distance between neighbouring cell. According to the first analogue proposal to quantum Hall effect~\cite{WangZ2008}, we choose for our pillars an anisotropic magnetic permeability 
\begin{equation}
\bar{\bar{\mu}}_{\rm d}=\left(\begin{array}{ccc}
\mu & i \kappa& 0  \\
-i\kappa & \mu& 0  \\
0 & 0 & 1
\end{array}\right), 
\end{equation}
under external d.c. magnetic field in $z$ direction, along with diagonal permeability $\mu=0.84$. For transverse magnetic (TM) mode (${E}_{z}$, ${H}_{x}$ and ${H}_{y}$ components only), the band degeneracy and its breaking at $\Gamma$ point are manipulated by tuning the rotation angle $\theta$ of hexagons in every unit. \Y{Here the TE mode is defined to indicate the eigenmode with electric polarization perpendicular to $xy$ plane as in wave equation $\nabla\times(\nabla\times E_z\hat{z})-k_0^2\epsilon_{\rm r}E_z\hat{z}=0$. } Furthermore we twist each unit cell with certain randomness by adding two random degrees of freedom: the variances to rotation angle $\theta$ and to pillar permittivity $\varepsilon_{\rm d}$, as indicated in Fig.~\ref{fig1}(c). 

For Chern number in a 2D periodic lattice, it can be defined as \cite{ZhaoR2020, Haixiao2020}: for TM mode of a 2D PhC, Chern number of the $n^{\rm th}$ band can be defined by integrating
\begin{eqnarray}
C^{(n)}&=&\frac{1}{2\pi}\int_{\rm BZ}\mathbf{F}_n(\mathbf{k})\cdot{\rm d}\mathbf{k}\\
&\approx&\frac{1}{2\pi}\sum_{\rm BZ}F_{\mathbf{k}}^{(n)}\Delta S_{\mathbf{k}}\\
\label{loop}&=&\frac{1}{2\pi}\sum_{\rm BZ}\Im \log[U^{(n)}_{\mathbf{k}_1\rightarrow\mathbf{k}_2}U^{(n)}_{\mathbf{k}_2\rightarrow\mathbf{k}_3}U^{(n)}_{\mathbf{k}_3\rightarrow\mathbf{k}_4}U^{(n)}_{\mathbf{k}_4\rightarrow\mathbf{k}_1}]. \nonumber \\
\end{eqnarray}
In the numeric integral method, symbol $\Im$ stands for the imaginary part, and $U^{(n)}_{\mathbf{k}_\alpha \rightarrow\mathbf{k}_\beta}:=\langle\mathbf{u}_{\mathbf{k}_\alpha}^{(n)}\vert \mathbf{u}_{\mathbf{k}_\beta}^{(n)}\rangle/\vert\langle\mathbf{u}_{\mathbf{k}_\alpha}^{(n)}\vert \mathbf{u}_{\mathbf{k}_\beta}^{(n)}\rangle\vert \quad (\alpha, \beta=1, 2, 3, 4)$ has been normalised from the inner product of eigenmodes $\mathbf{u}_{\mathbf{k}_{\alpha,\beta}}$, which with the permittivity weight integrates as 
\begin{equation}
\langle\mathbf{u}_{\mathbf{k}_\alpha}^{(n)}\vert \mathbf{u}_{\mathbf{k}_\beta}^{(n)}\rangle:=\iint\limits_{\rm unit\quad cell} {\rm d}x{\rm d}y \epsilon(\mathbf{r})\mathbf{u}_{\mathbf{k}_\alpha}^*\cdot \mathbf{u}_{\mathbf{k}_\beta}. 
\end{equation}
With sufficient meshes taken in the first Brillouin zone, the four multiplication loop in Eq.~\eqref{loop} shall give accurate values for Chern numbers~\cite{Fukui2005, JinD2017}. \Y{And our topological phases are completely pinpointed by the integer number of computed Chern number.} In Sec.~\ref{sec3} this numeric method is adopted to calculate the numeric values of Chern numbers for PhC under periodic perturbation. Under such periodic perturbation, the numeric method to calculate the Chern number still applies for the periodic condition is kept. \Y{Note that in our paper we do not consider the dielectric loss of material. When this is considered, edge states would travel less long due to the loss and are worth further investigation beyond this paper. }

\begin{figure*}[htbp]
\includegraphics[width=0.5\textwidth]{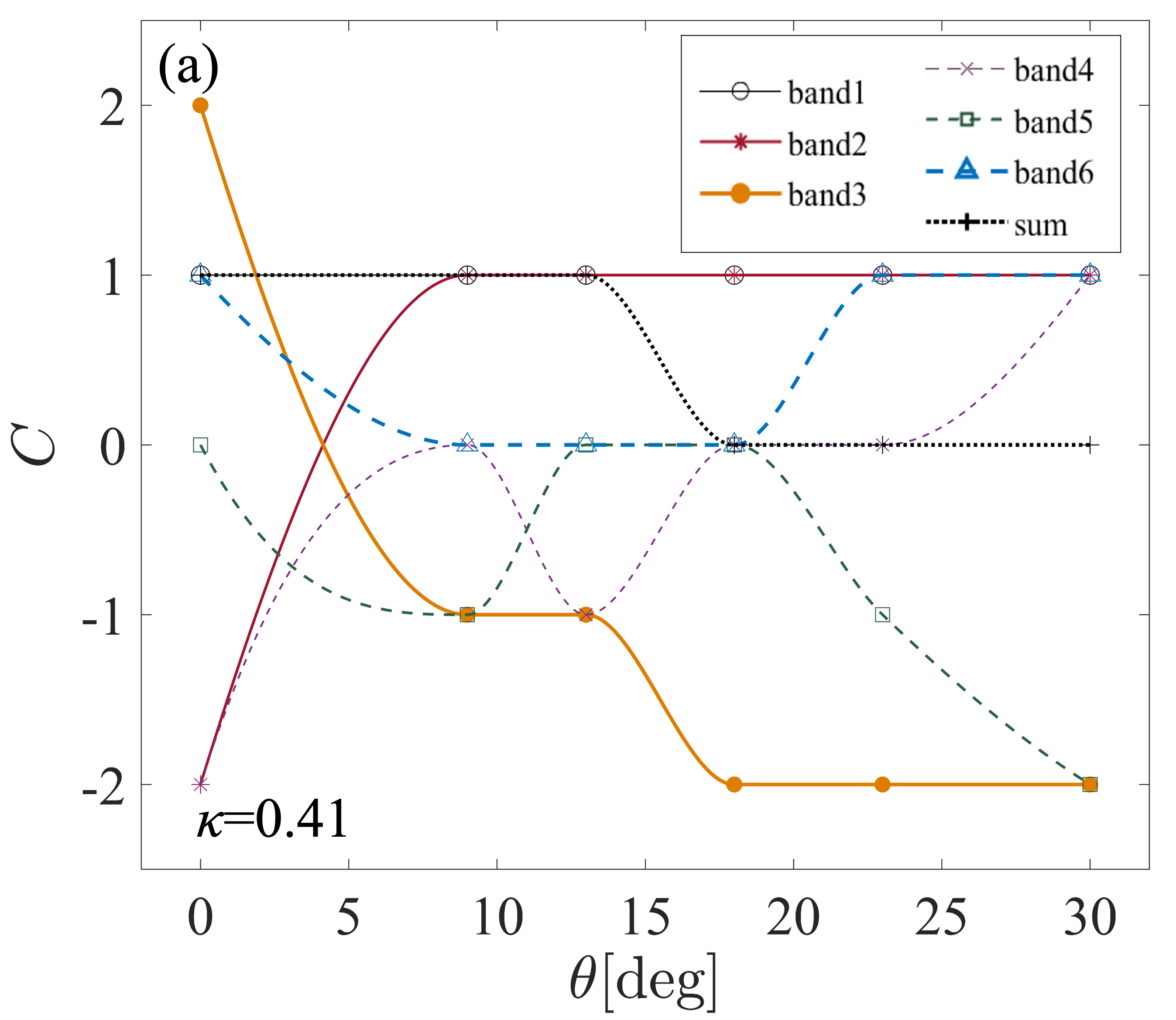}\\
\includegraphics[width=0.5\textwidth]{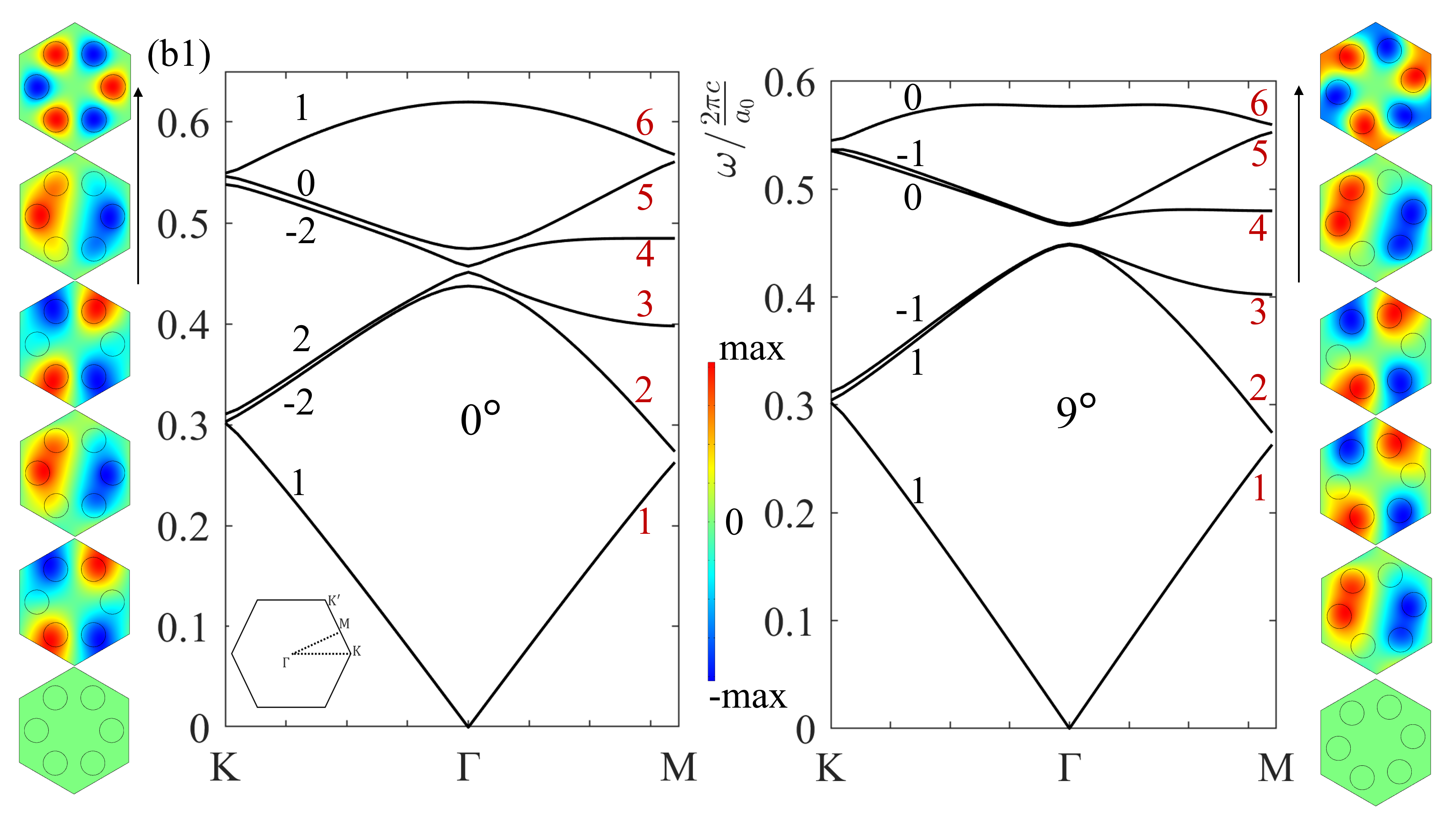}
\includegraphics[width=.5\textwidth]{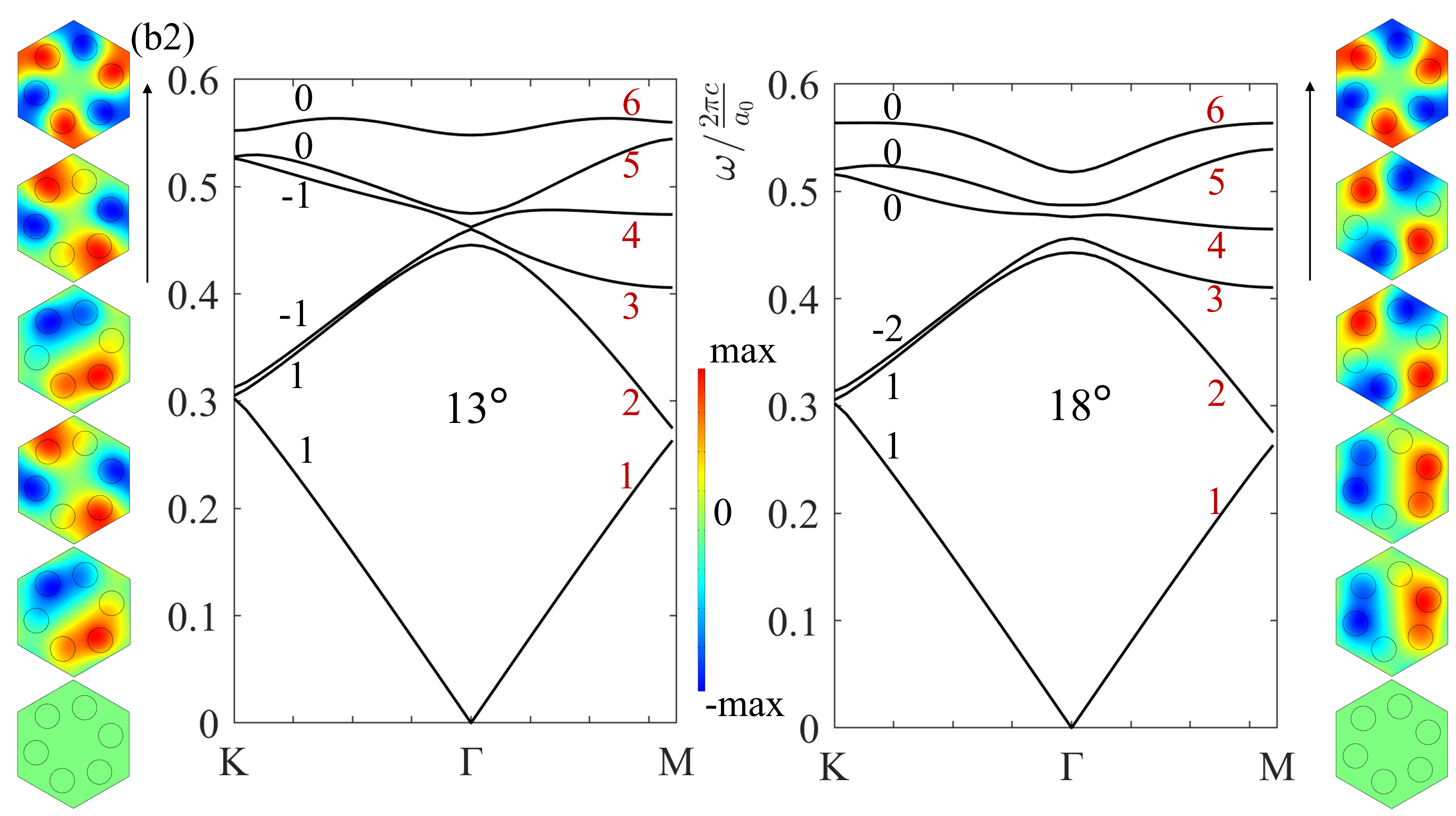}
\includegraphics[width=0.5\textwidth]{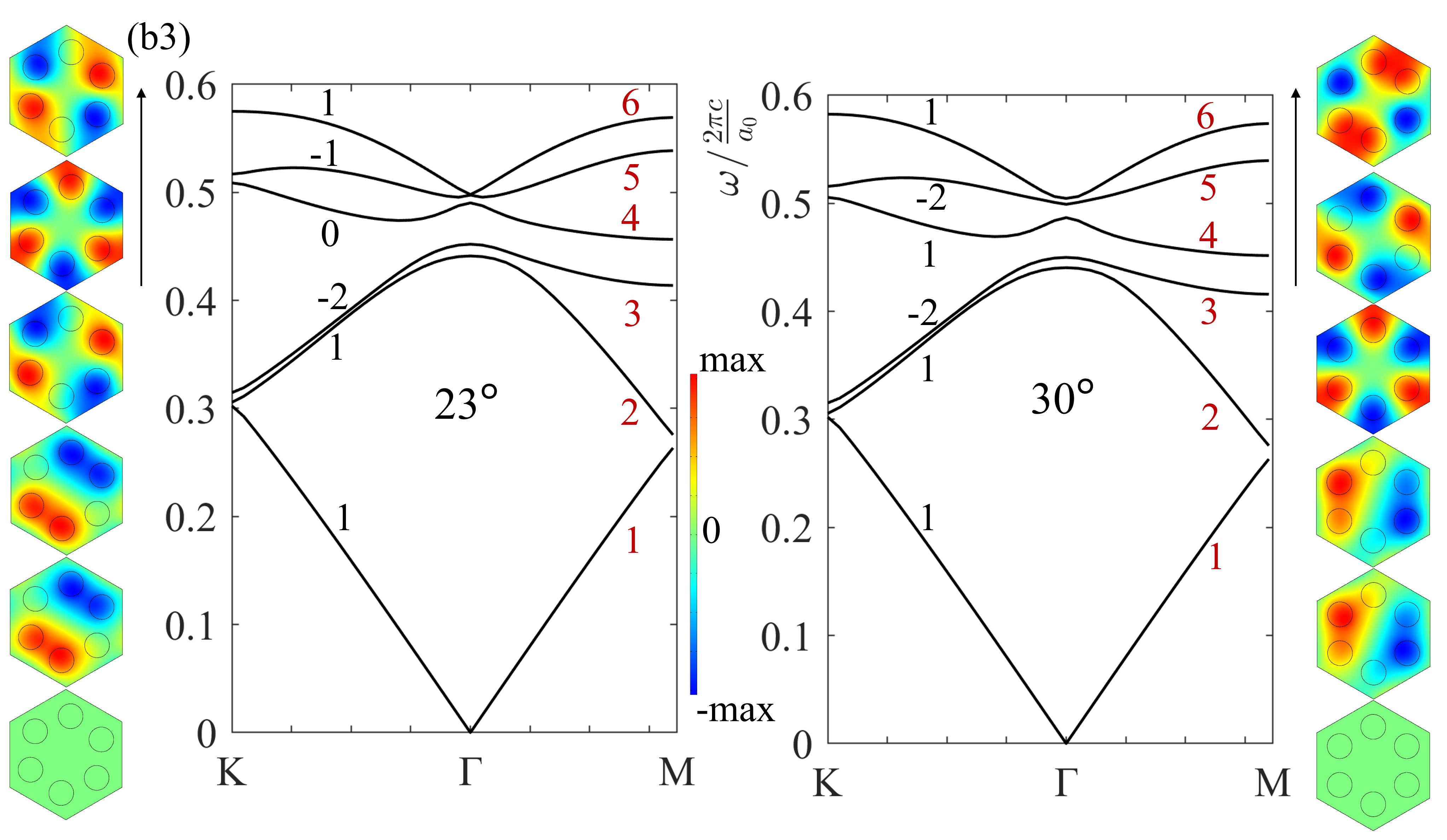}

\caption{\label{fig2} (a) Chern numbers versus rotation angle, with fitted curves using an interpolation method of piecewise cubic Hermite interpolation(PCHIP). (b) Band diagrams for six rotational angles which also embed Chern numbers with each band curve. Insets besides: electric fields of each eigenstate at $\Gamma$ points. Parameters: $\mu=0.84, \kappa=0.41$.  }
\end{figure*}

First for a multimode eigenstate~\cite{Skirlo2014, Skirlo2015} we choose the parameter of gyromagnetic pillar as $\mu=0.84, \kappa=0.41$. The Chern numbers of all six bands, including the sum of first three bands, are calculated according to Eq.~\eqref{loop}, shown in Fig.~\ref{fig2}. Here in our paper we consider the edge states in the gap above the first three bands. We find that as the rotation angle varies from $0^\circ$ to $30^\circ$, the sum of the first three bands reduces from 1 to 0 since $\theta>15^\circ$, and the respective Chern number for each band switches mosaically with no apparent guideline, which are all plotted in panel (a-b). We note that Chern numbers should be integers and our curves linking all integer Chern numbers are just to guide the eyes to follow the phase varying trend. The eigenstates for each band at $\Gamma$ point are also presented as insets of panel (b). It is curious that Chern number of band 3 changes abruptly from 2 to -2 in our data range. 
 
\begin{figure}
\includegraphics[width=0.5\textwidth]{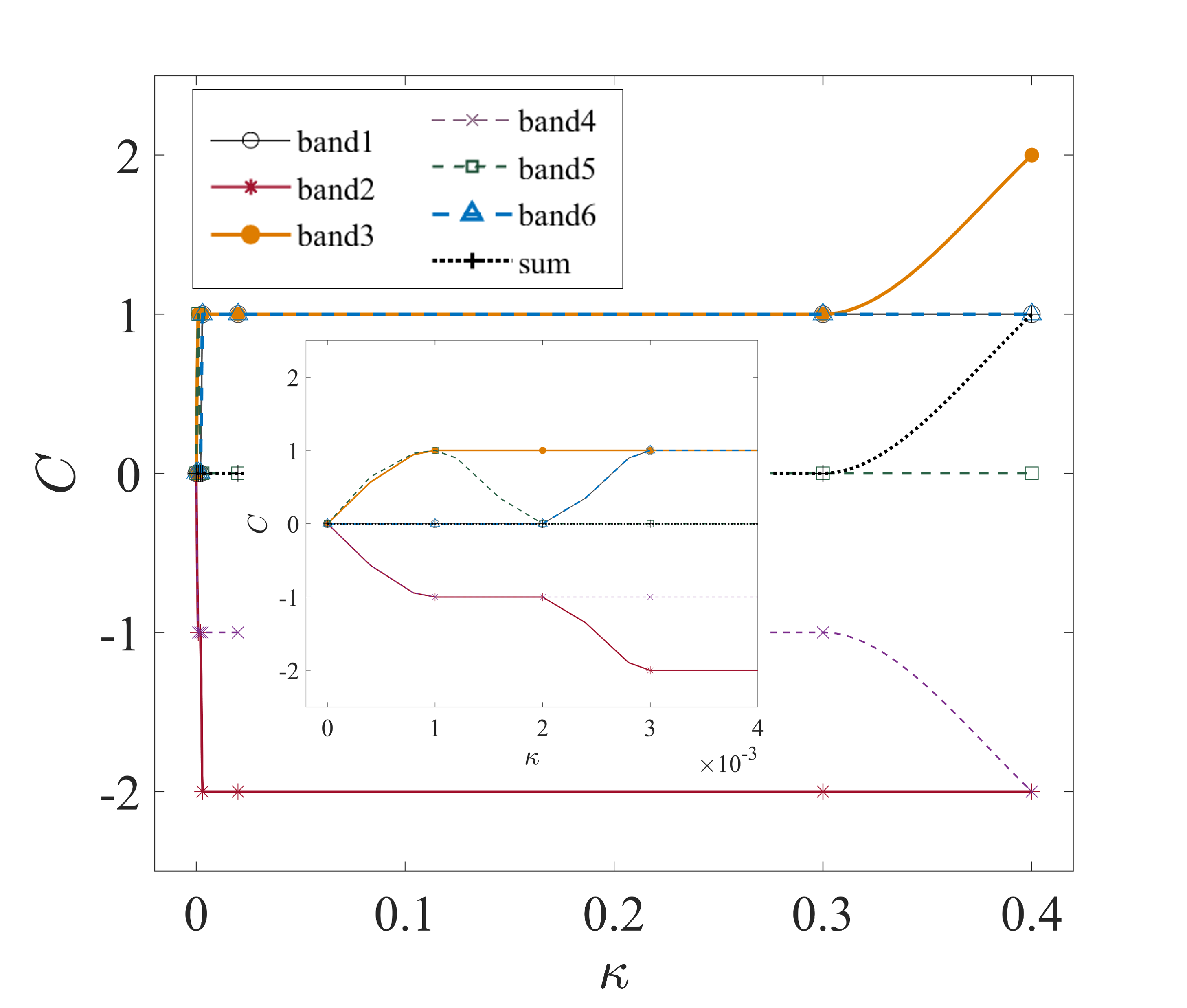}
\caption{\label{fig3} Chern numbers for band 1-6 along with the sum for the first three bands, versus gyromagnetic coefficient $\kappa$. Inset: zoom up for $\kappa<0.004$.  Parameters: $\mu=0.84, \theta =0^\circ$. }
\end{figure}

Since no clear guidance to tune Chern numbers were found for such large gyromagnetic parameter of $\kappa =0.41$, we tune the off-diagonal permeability $\kappa$ to smaller range to find out its least possible value to tune Chern numbers of our model. In Fig.~\ref{fig3}, the Chern numbers of all six bands including the sum for the first three are presented with varying the off-diagonal permeability in a relatively small range $\kappa \leq0.40$. No clear guidance is found to relate the bands and Chern number within available data. Among them, large Chern numbers of two are also reached for band 2 at a small value $\kappa<0.004$, and for band 3 only at a large value $\kappa=0.4$. 

\begin{figure*}
\includegraphics[width=0.5\textwidth]{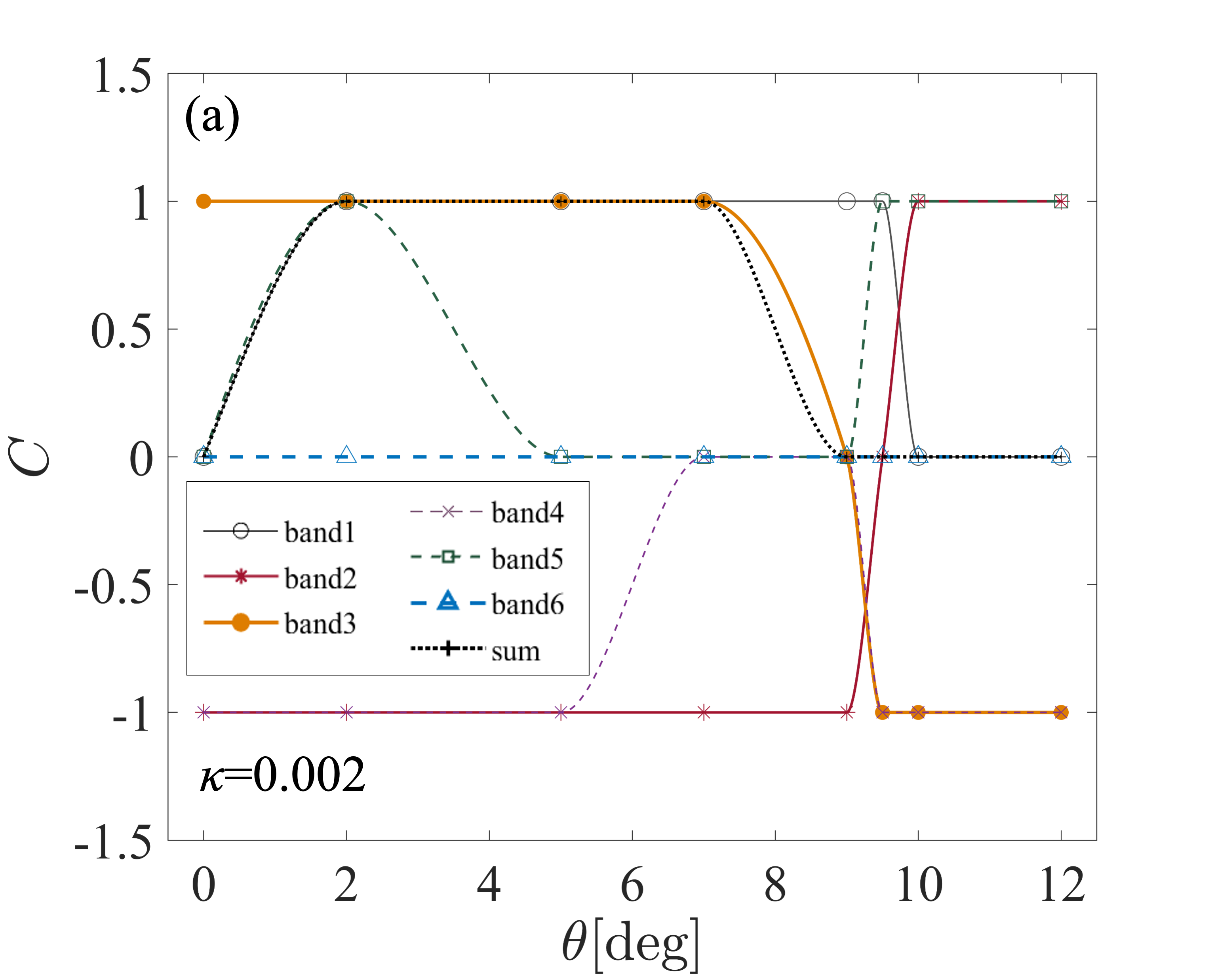}
\includegraphics[width=0.75\textwidth]{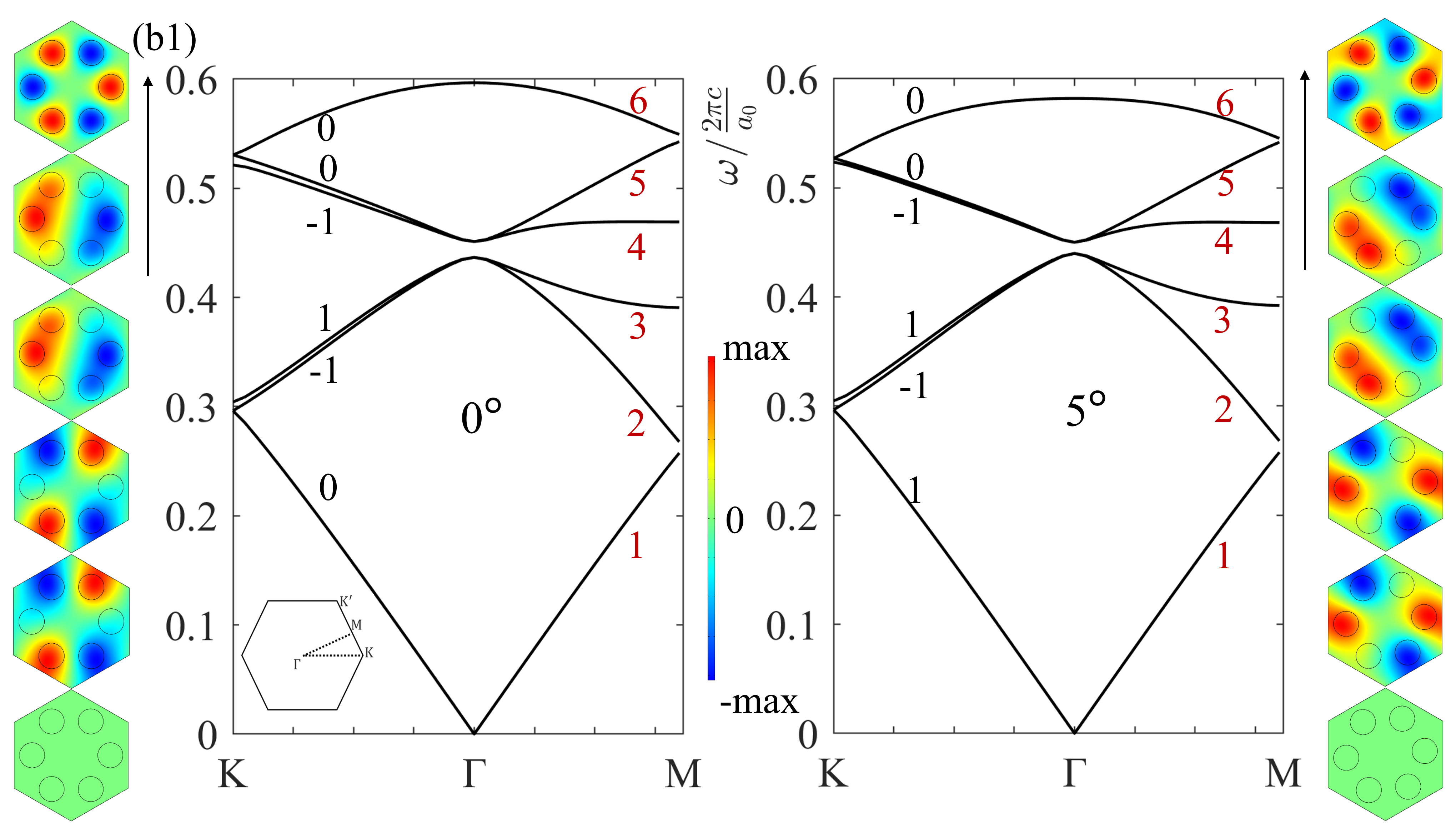}
\includegraphics[width=0.75\textwidth]{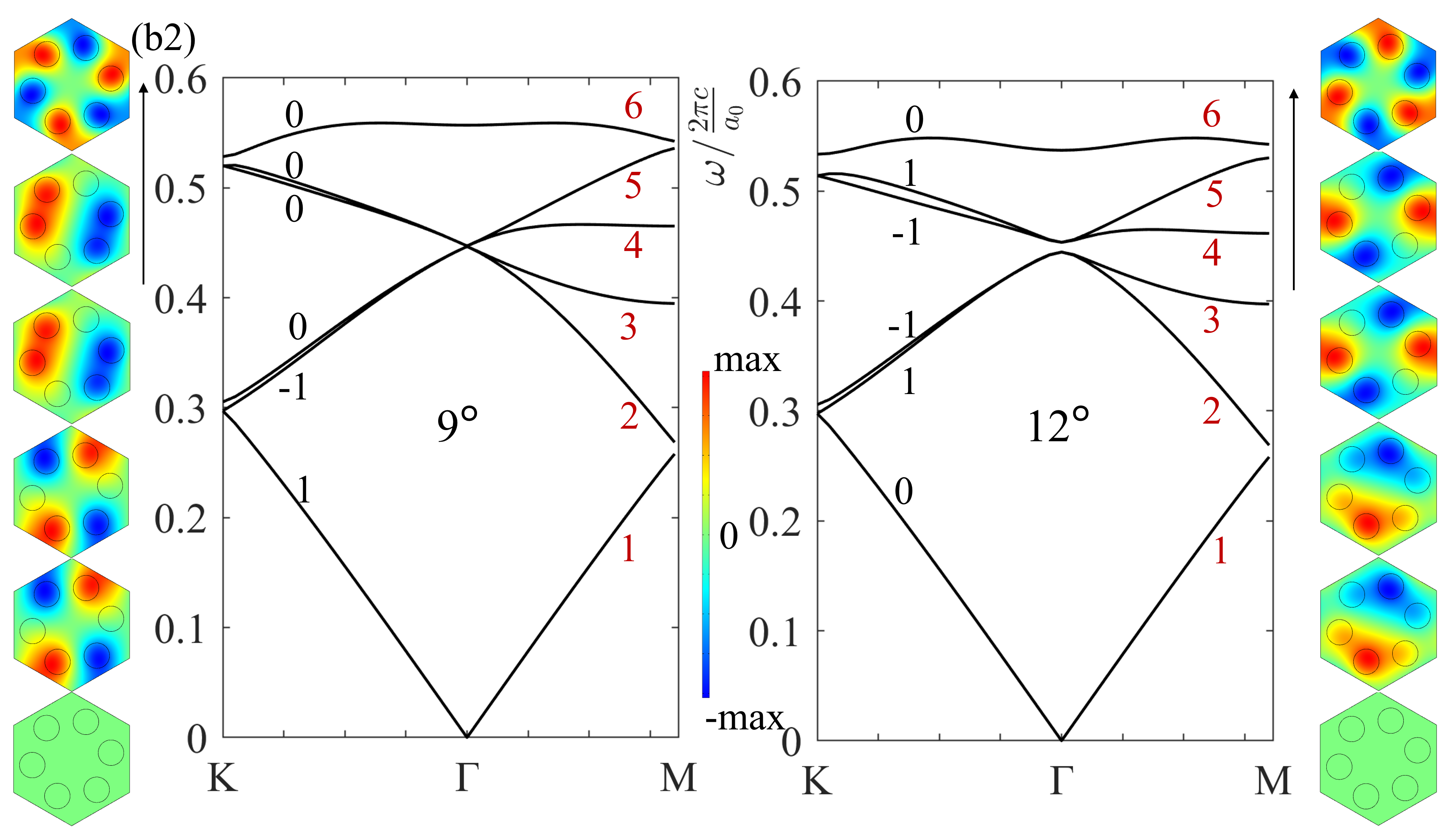}

\caption{\label{fig4} Chern numbers for band 1-6  with the sum for the first three bands, versus the rotation angle $\theta$ for $\mu=0.84, \kappa=0.002$. }
\end{figure*}

We choose the off-diagonal permeability $\kappa=0.002$ as a reference parameter in order to find out the disruption effect to Chern numbers of chiral edge states under periodic perturbation. For regular lattice without disorder, Chern numbers for all six bands and the sum of the first three bands are calculated and presented in Fig.~\ref{fig4}(a). Also the band diagrams for four rotation angles are presented in Fig.~\ref{fig4}(b1-b2), along with their electric field amplitudes of eigenstates at $\Gamma$. Curiously, the Chern-number sum of the first three bands evolves as the rotation angle $\theta$ changes from 0 to $12^\circ$, which changes from 0 to 1 at around $\theta=2^\circ$ and falls back to 0 at around $\theta=2^\circ$. This demonstrates the vulnerableness and sensitiveness of Chern number under the rotation angle changes. As next Sec.~\ref{sec3} will show, when periodic disruption pushes in, Chern numbers evolve from 1 back to 0 in a more vulnerable manner.

\section{\label{sec3} Disordered lattice: disruption to Chern numbers}
In this Sec., we will calculate Chern numbers for our model under periodic disruption to investigate the breakdown effect for topological invariants in our PhC model. Specifically, we choose two random parameters: the variances to rotation angle $\mu_1:=\theta$ and to pillar permittivity $\mu_2:=\varepsilon_{\rm d}$ under random perturbation. For every random ensemble,  Chern numbers of 10 randomly-chosen samples are calculated and processed. 

\begin{figure*}[!htbp]
\includegraphics[width=0.49\textwidth]{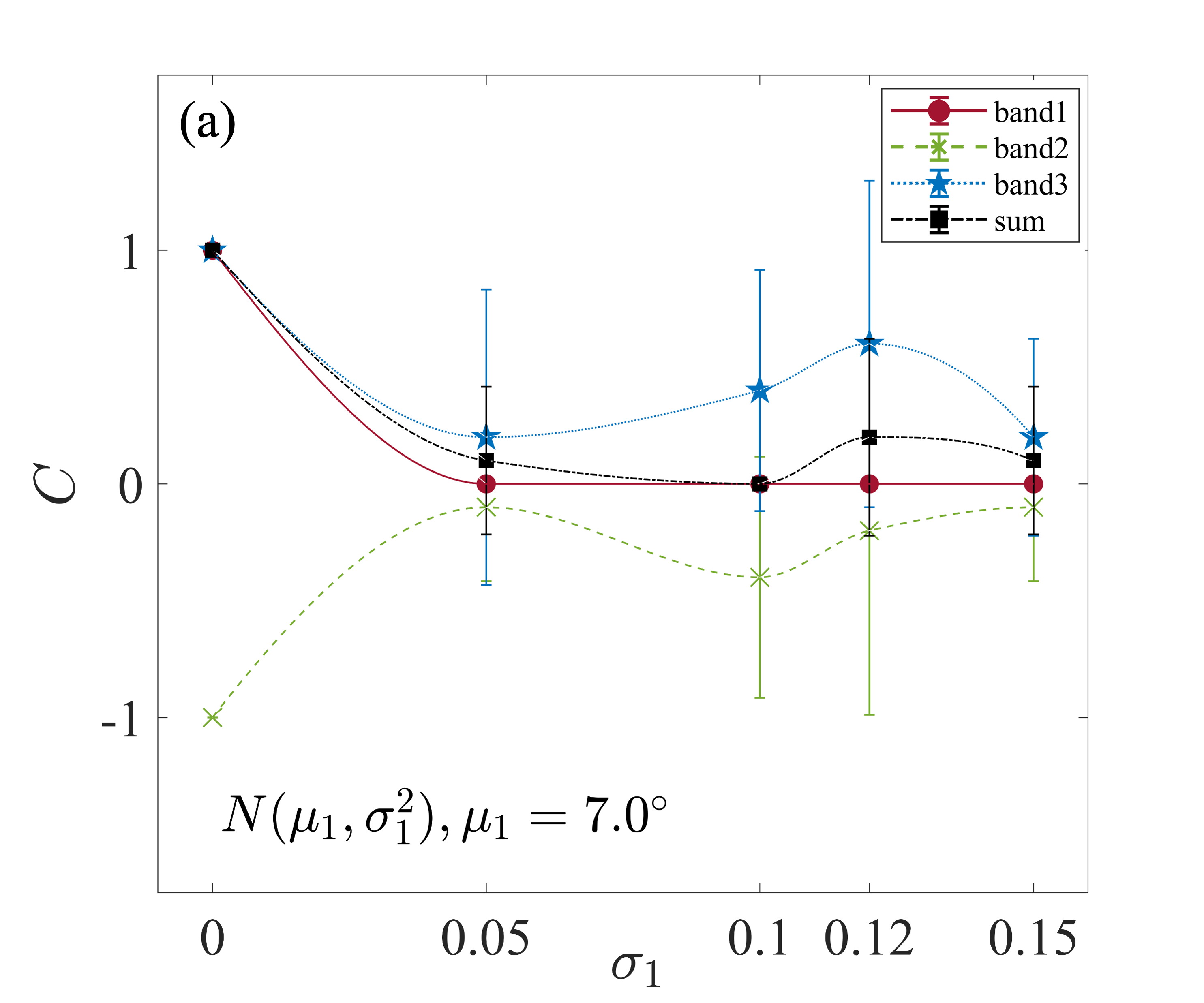}
\includegraphics[width=0.49\textwidth]{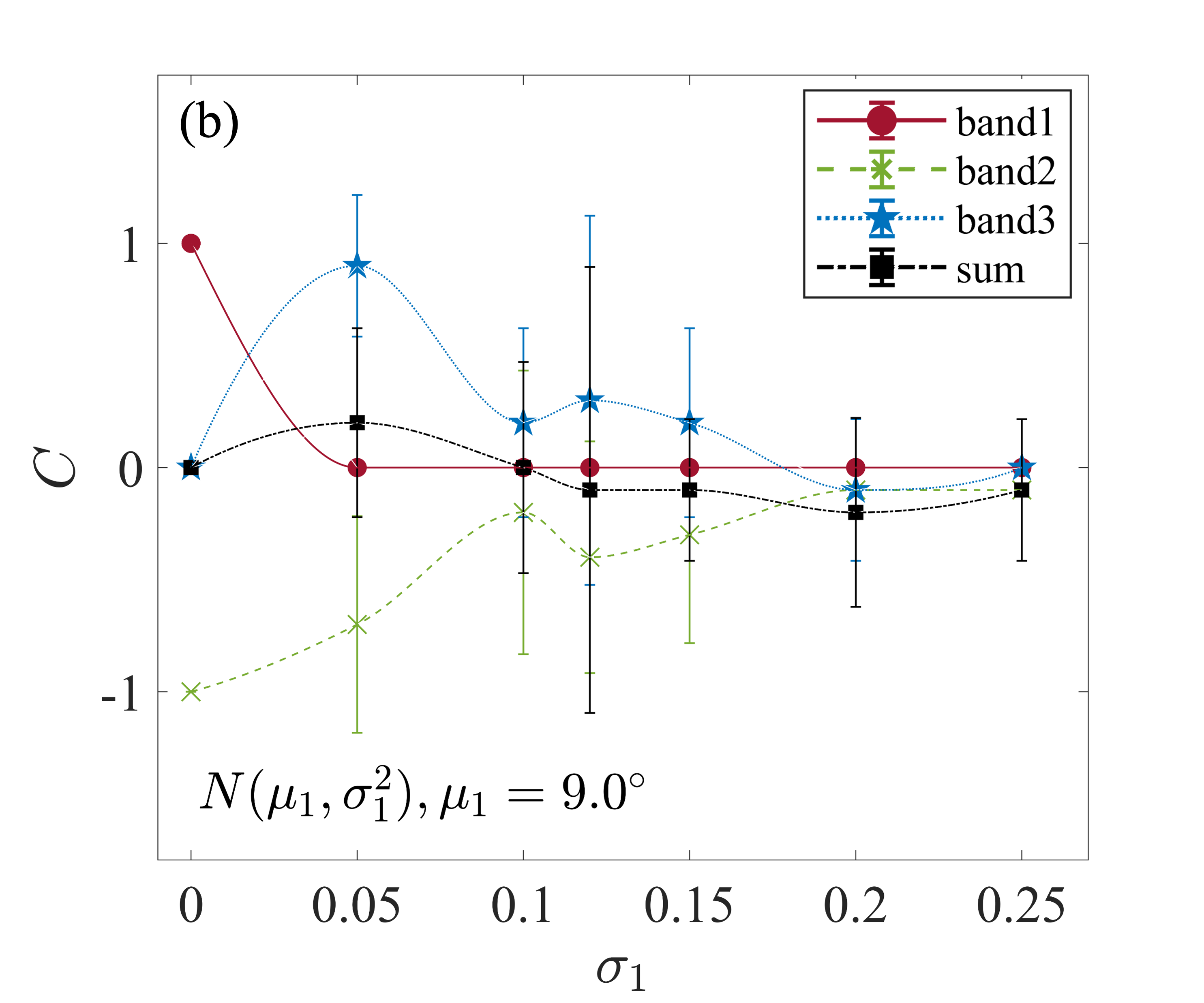}\\
\includegraphics[width=0.49\textwidth]{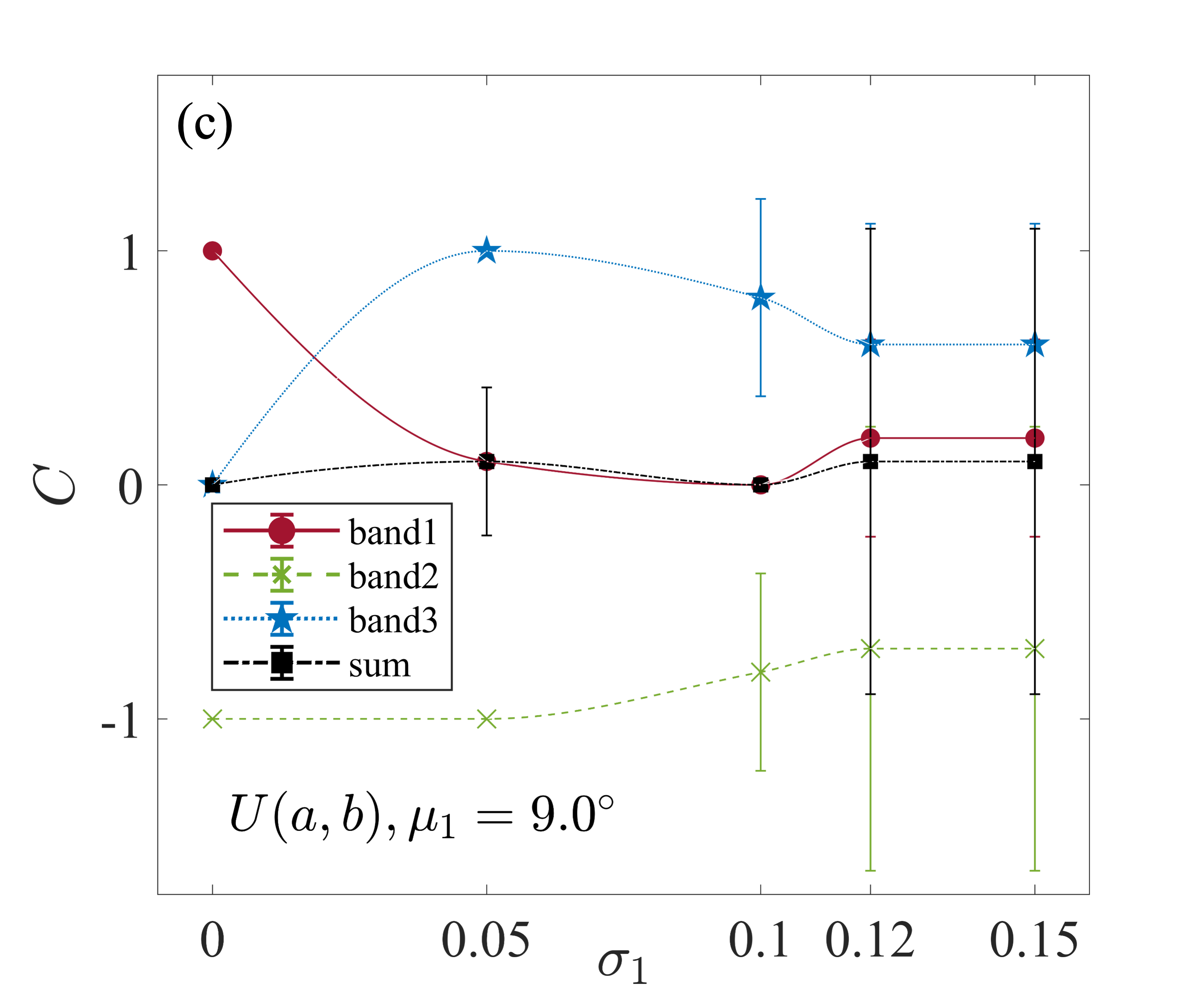}
\includegraphics[width=0.49\textwidth]{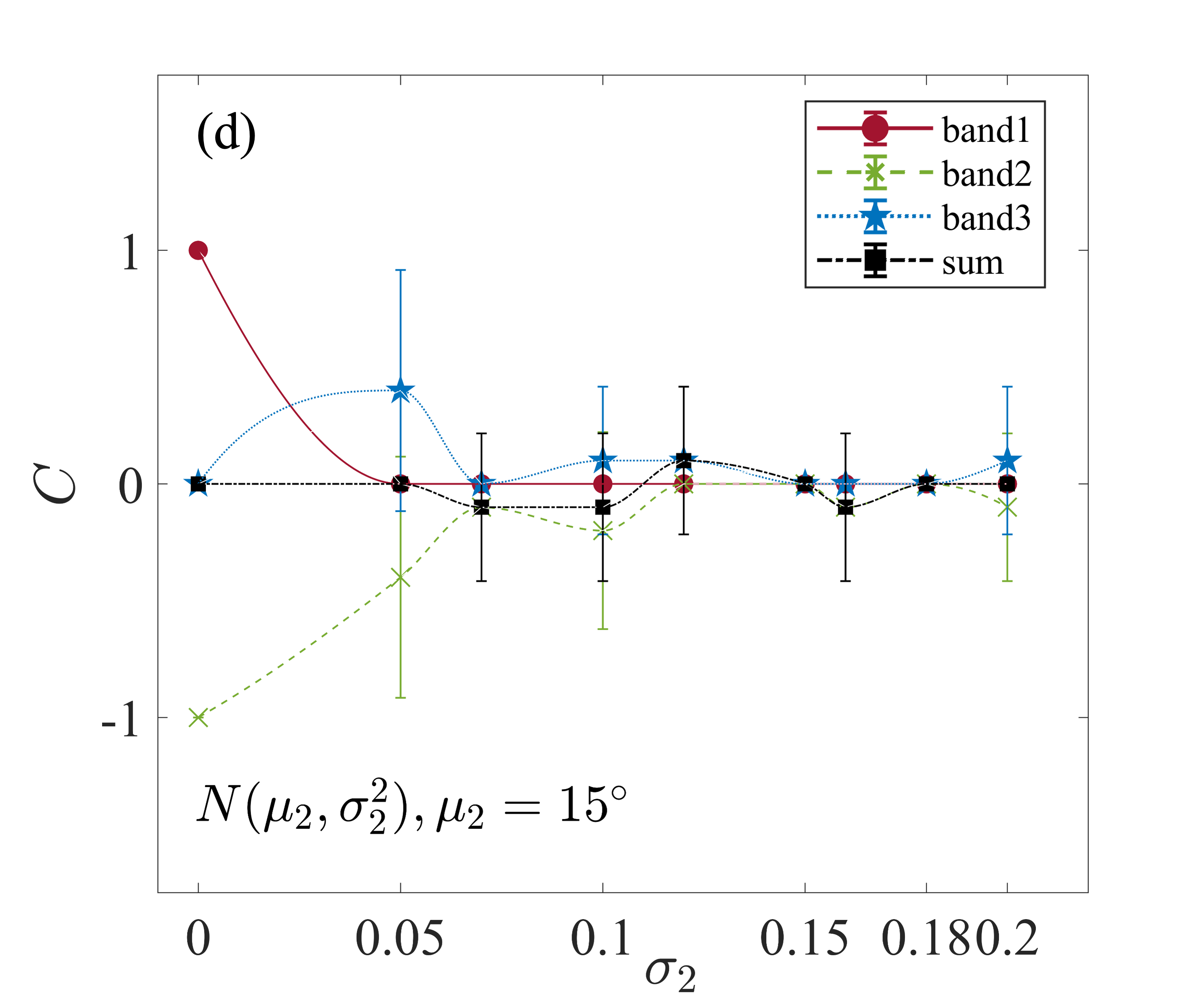}
\caption{\label{fig5} Chern numbers for the first three bands and their sum versus standard deviation $\sigma_{1,2}$ for (a-b) rotational angles under Gaussian randomization respectively (a) $\mu_1=7.0^\circ$, (b) $\mu_1=9.0^\circ$ and under uniform distribution (c) $\mu_1=9.0^\circ$; (d) Pillar permittivity under Gaussian randomization $\mu_2=15$.}
\end{figure*}

Fig.~\ref{fig5}(a-d) contains four sets of calculated Chern numbers for four types of disordered PhC under periodic perturbation, which exerts on rotation angle and pillar material of every unit cell. Two kinds of random distributions are chosen to illustrate the break-down effect of topological phases: (a, b, d) for Gaussian distribution $\mathcal{N}(\mu_1, \sigma_1)$ and (c) $\mathcal{U}(\mu_1, \sigma_1)$.  Chern numbers are plotted when Gaussian distribution and uniform distribution are introduced to rotational angle $\mu_1=\theta$ [\emph{cf.} panels (a-c)] and pillar permittivity $\mu_2=\varepsilon_{\rm d}$ [\emph{cf.} panel (d)].

We observe that (1) band 3 occupies robust non-trivial topological numbers against disorder, distinctively among all bands 1-3, which are presented in blue asterisks in Fig.~\ref{fig5}. On the other hand, band 2 also occupies certain non-trivial topological numbers probably due to the uniform distribution, which spreads more uniformly so that giving less disorder effectively. The contrast between the two random distribution implies that Gaussian distribution should give rise to severe symmetry breaking due to their narrowly spreading feature. 
(2) However, such stable nontrivial numbers add up still to zero for the first three bands. In a word, the sums of topological numbers for the first three bands, marked in black squares, all reduce to zero when standard deviations $\sigma_{1, 2}$ are tuned large enough. Especially, some sums converge to stable values of zero, giving almost zero deviation, for example in certain data points of panels (a, c, d). 
(3) Panel (d) compared with panels (a-c) shows that unit perturbation on pillar permittivity $\varepsilon_{\rm d}$ are destructed more easily than on rotation angle $\theta$, which reveals a difference for permittivity parameter under periodic perturbation.  

As Fig.~\ref{fig5} demonstrate, all Chern numbers degrade to zero for sufficiently large standard deviation $\sigma_1, \sigma_2$. Moreover, the sum of Chern numbers of the first three bands also degrades quickly when the unit cells randomizes further enough. Regardlessly one sees that Chern number for a band in an occasional sample, could be tuned from zero to nonzero~[data not shown here, cf.~\cite{TianThesis2023}] with the apparent impression that topological Anderson insulator \emph{could} still occur there. However, after simulating with a sufficient sample number, this turns out \emph{not} the case. Some data points in Fig.~\ref{fig5} reflect this point, which give stable non-integer average numbers, for instance, a few symbols for band 3. However, they do not indicate non-trivial topological phases because topological numbers should be integers according to the definition. Such fractional numbers in statistical average only suggests the breakdown effect of topological phase of all samples in an ensemble under periodic perturbation.


\begin{figure*}[!htbp]
\includegraphics[width=0.93\textwidth]{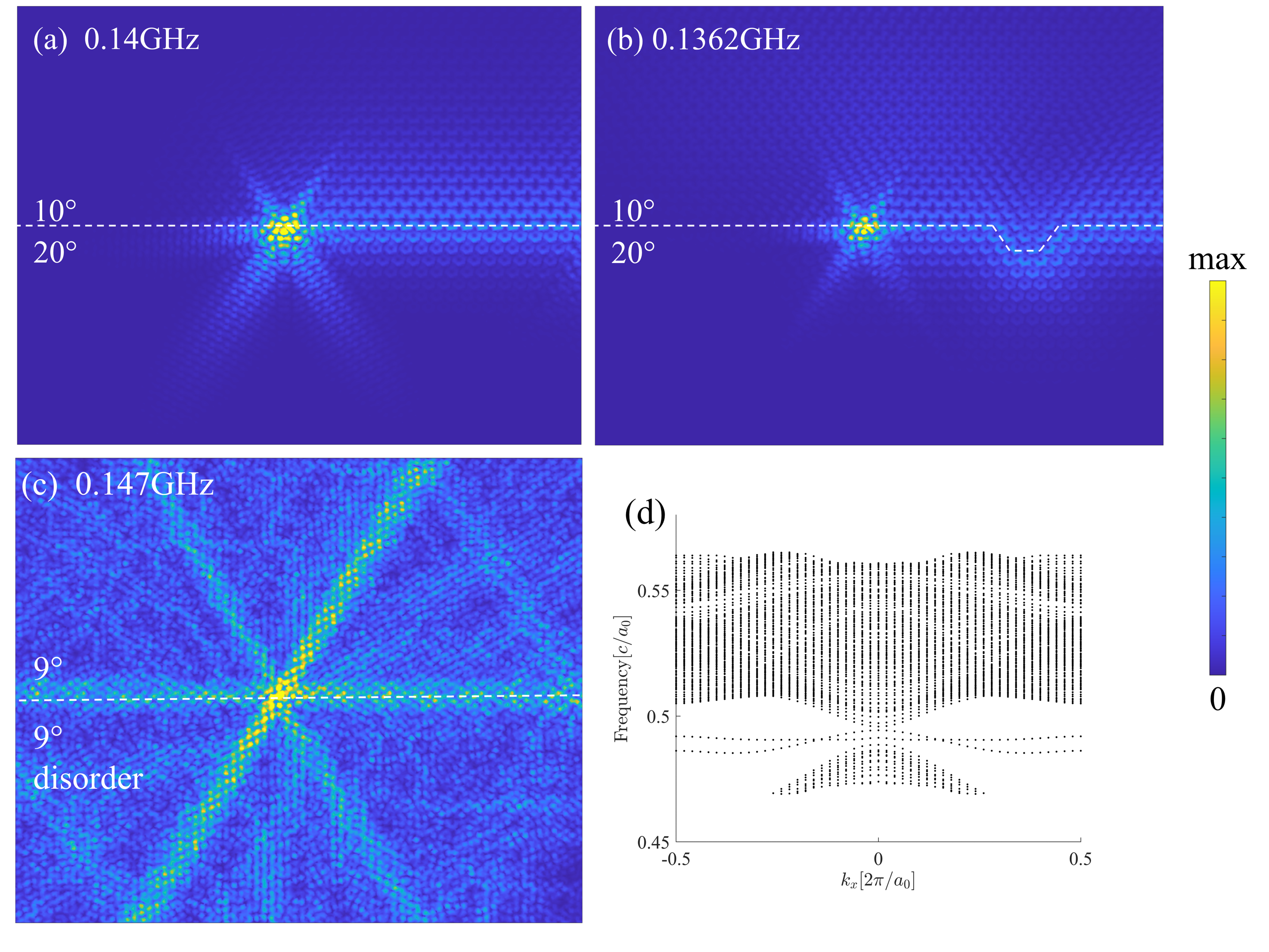}

\caption{\label{fig6} 
(a) Electric field for a topological edge states between two gyromagnetic PhC with different rotational angles; (b) Electric field for a topological edge states between two gyromagnetic PhC with different rotational angles under a disruption for several units; (c) Electric field for an edge state positioned sandwiched between two gyromagnetic PhCs of unit cells rotated by the same angle $\mu_1$: one for a periodic lattice and the other a disordered one under Gaussian randomization $\mathcal{N}(\mu_1 = 9^\circ, \sigma_1 = 0.12)$. Source frequency: (a) $ 0.140{\rm GHz}$, (b) $0.136{\rm GHz}$, (c) $ 0.147{\rm GHz}$. (d) Dispersion diagram for (c) edge states. }

\end{figure*}

To illustrate this point of trivial topological numbers, we demonstrate the electric field for a gap mode to check its robust edge states in Fig.~\ref{fig6}. A wave source is put at a certain point along the edge between two types of PhC. Before adding randomization, we create an interface between two structures with different rotation angles thereby having different Chern numbers, and demonstrate a topological edge mode in Fig. 6(a). Meanwhile we demonstrate a topological edge mode through a local rotation of several unit cells in Fig. 6(b) in the same sprit of~\cite{WangX2020, ChenJ2022}. Nevertheless, in panel (c) randomization comes into play: we make one regular lattice and the other disordered by Gaussian distribution $\mathcal{N}(\mu_1=9^\circ, \sigma_1=0.12)$ (see sampled angles in Tab.~\ref{table1}), both with rotation angle $\theta=9^\circ$. Panel (c) shows an edge state as expected and also some other chaotic emission in unexpected directions. And the projected band diagram in panel (d) indicates two tangled in-gap edge dispersion curves close to emission frequency $0.147{\rm GHz}$. 

As to the reason why topological phase breaks down easily by the periodic perturbation, it is speculated that such defined disorder breaks the $\mathcal{C}_6$ spatial symmetry, thus ruling out the non-trivial interaction in unit cells which emerges into topological exotic phases~\cite{ZhouR2021}. 

Therefore we sum up Sec.~\ref{sec3} that, in our gyromagnetic PhC, topological phases induced from synchronized rotation are fragile and easily break down due to periodic perturbation on rotation angle $\theta$ and to pillar permittivity $\varepsilon_{\rm d}$. And we attribute its reason to the broken symmetry resultant from certain disorder.

\begin{table}[!htbp]
\caption{\label{table1} Rotational angles for periodic perturbation for PhC in Fig.~\ref{fig6}, which are a sample following Gaussian distribution $\mathcal{N}(\mu_1=9.0^\circ, \sigma_1=0.12)$. }
\begin{ruledtabular}
\begin{tabular}{llllll}
$\theta_1$ &$ \theta_2$ &$\theta_3$&$\theta_4$&$\theta_5$&$\theta_6$\\
$9.0046^\circ$ & $9.0223^\circ$ & $9.0020^\circ$ & $9.0324^\circ$& $9.0593^\circ$ & $8.9715^\circ$\\
\end{tabular}
\end{ruledtabular}
\end{table}

\section{\label{sec4}Conclusion}

In conclusion, we calculate the topological invariants for every band in a gyromagnetic lattice with synchronized rotation of unit cells to induce topological phase transition. Based on that we investigate the variance on topological invariants for our periodic Kekul\'e lattice under periodic perturbation by randomizing two structural parameters, one for rotational angle and the other pillar permittivity. Our numeric results reveal the vulnerability of topological phases which are easily broken down by geometric and material disorders in such wave-analogue systems. This work provides access to check robustness of topological edge states under such periodic unit-perturbation in pragmatic experiments and further progress on breakdown effect by other randomly-disrupted lattice is underway. 

Note to add: we find a recent paper~\cite{ZhangH2022} on disorder-driven collapse of topological phases after submission of our paper.


\emph{Author statement.}
Y. Tian: Software, Validation, Formal analysis, Investigation, Data curation, Visualization, Writing - Original Draft. 
R. Zhou: Methodology, Software, Validation, Visualization. 
Z. Liu: Methodology, Software, Writing - Review \& Editing. 
Y. Liu: Conceptualization, Methodology, Investigation, Resources, Writing - Original Draft, Visualization, Supervision, Project administration, Funding acquisition, Writing - Review \& Editing. 
H. Lin: Resources, Project administration. 
B. Zhou: Conceptualization, Funding acquisition.


\begin{acknowledgments}

Y. T. and Y. L. thank Xu Dong-Hui, Wang Haixiao, Chen Zhaoxian, Chen Huan and Chen Rui for helpful discussion, and are supported by Young Scientist Fund [NSFC11804087], National Natural Science Foundation [NSFC12247101]; Science and Technology Department of Hubei Province [2022CFB553]; Hubei University [X202210512039]; and Educational Commission of Hubei Province of China [Q20211008]. R. Z. and H. L. are supported by the fundamental Research Funds for
the Central University of China [CCNU19TS073]. Z.-R. L., Y. L. and B. Z. is supported by National Natural Science Foundation of China [NSFC12074107], Science and Technology Department of Hubei Province [2022CFA012], and Educational Commission of Hubei Province [T2020001]. 


\end{acknowledgments}
\emph{Declaration of interests.} The authors declare that they have no known competing financial interests or personal relationships that could have appeared to influence the work reported in this paper.


\bibliography{ref6}

\providecommand{\noopsort}[1]{}\providecommand{\singleletter}[1]{#1}%
\begin{thebibliography}{44}%
\makeatletter
\providecommand \@ifxundefined [1]{%
 \@ifx{#1\undefined}
}%
\providecommand \@ifnum [1]{%
 \ifnum #1\expandafter \@firstoftwo
 \else \expandafter \@secondoftwo
 \fi
}%
\providecommand \@ifx [1]{%
 \ifx #1\expandafter \@firstoftwo
 \else \expandafter \@secondoftwo
 \fi
}%
\providecommand \natexlab [1]{#1}%
\providecommand \enquote  [1]{``#1''}%
\providecommand \bibnamefont  [1]{#1}%
\providecommand \bibfnamefont [1]{#1}%
\providecommand \citenamefont [1]{#1}%
\providecommand \href@noop [0]{\@secondoftwo}%
\providecommand \href [0]{\begingroup \@sanitize@url \@href}%
\providecommand \@href[1]{\@@startlink{#1}\@@href}%
\providecommand \@@href[1]{\endgroup#1\@@endlink}%
\providecommand \@sanitize@url [0]{\catcode `\\12\catcode `\$12\catcode
  `\&12\catcode `\#12\catcode `\^12\catcode `\_12\catcode `\%12\relax}%
\providecommand \@@startlink[1]{}%
\providecommand \@@endlink[0]{}%
\providecommand \url  [0]{\begingroup\@sanitize@url \@url }%
\providecommand \@url [1]{\endgroup\@href {#1}{\urlprefix }}%
\providecommand \urlprefix  [0]{URL }%
\providecommand \Eprint [0]{\href }%
\providecommand \doibase [0]{https://doi.org/}%
\providecommand \selectlanguage [0]{\@gobble}%
\providecommand \bibinfo  [0]{\@secondoftwo}%
\providecommand \bibfield  [0]{\@secondoftwo}%
\providecommand \translation [1]{[#1]}%
\providecommand \BibitemOpen [0]{}%
\providecommand \bibitemStop [0]{}%
\providecommand \bibitemNoStop [0]{.\EOS\space}%
\providecommand \EOS [0]{\spacefactor3000\relax}%
\providecommand \BibitemShut  [1]{\csname bibitem#1\endcsname}%
\let\auto@bib@innerbib\@empty
\bibitem [{\citenamefont {Kane}\ and\ \citenamefont {Mele}(2005)}]{Kane2005a}%
  \BibitemOpen
  \bibfield  {author} {\bibinfo {author} {\bibfnamefont {C.~L.}\ \bibnamefont
  {Kane}}\ and\ \bibinfo {author} {\bibfnamefont {E.~J.}\ \bibnamefont
  {Mele}},\ }\bibfield  {title} {\bibinfo {title} {Quantum spin hall effect in
  graphene},\ }\href@noop {} {\bibfield  {journal} {\bibinfo  {journal}
  {Physical Review Letters}\ }\textbf {\bibinfo {volume} {95}},\ \bibinfo
  {pages} {226801} (\bibinfo {year} {2005})}\BibitemShut {NoStop}%
\bibitem [{\citenamefont {Haldane}\ and\ \citenamefont
  {Raghu}(2008)}]{Haldane2008}%
  \BibitemOpen
  \bibfield  {author} {\bibinfo {author} {\bibfnamefont {F.~D.~M.}\
  \bibnamefont {Haldane}}\ and\ \bibinfo {author} {\bibfnamefont
  {S.}~\bibnamefont {Raghu}},\ }\bibfield  {title} {\bibinfo {title} {Possible
  realization of directional optical waveguides in photonic crystals with
  broken time-reversal symmetry},\ }\href
  {https://doi.org/10.1103/PhysRevLett.100.013904} {\bibfield  {journal}
  {\bibinfo  {journal} {Phys. Rev. Lett.}\ }\textbf {\bibinfo {volume} {100}},\
  \bibinfo {pages} {013904} (\bibinfo {year} {2008})}\BibitemShut {NoStop}%
\bibitem [{\citenamefont {Wang}\ \emph {et~al.}(2008)\citenamefont {Wang},
  \citenamefont {Chong}, \citenamefont {Joannopoulos},\ and\ \citenamefont
  {Solja{\v c}i{\'c}}}]{WangZ2008}%
  \BibitemOpen
  \bibfield  {author} {\bibinfo {author} {\bibfnamefont {Z.}~\bibnamefont
  {Wang}}, \bibinfo {author} {\bibfnamefont {Y.~D.}\ \bibnamefont {Chong}},
  \bibinfo {author} {\bibfnamefont {J.~D.}\ \bibnamefont {Joannopoulos}},\ and\
  \bibinfo {author} {\bibfnamefont {M.}~\bibnamefont {Solja{\v c}i{\'c}}},\
  }\bibfield  {title} {\bibinfo {title} {Reflection-free one-way edge modes in
  a gyromagnetic photonic crystal},\ }\href@noop {} {\bibfield  {journal}
  {\bibinfo  {journal} {Physical Review Letters}\ }\textbf {\bibinfo {volume}
  {100}},\ \bibinfo {pages} {013905} (\bibinfo {year} {2008})}\BibitemShut
  {NoStop}%
\bibitem [{\citenamefont {Hasan}\ and\ \citenamefont {Kane}(2010)}]{Hasan2010}%
  \BibitemOpen
  \bibfield  {author} {\bibinfo {author} {\bibfnamefont {M.~Z.}\ \bibnamefont
  {Hasan}}\ and\ \bibinfo {author} {\bibfnamefont {C.~L.}\ \bibnamefont
  {Kane}},\ }\bibfield  {title} {\bibinfo {title} {Colloquium: topological
  insulators},\ }\href@noop {} {\bibfield  {journal} {\bibinfo  {journal}
  {Reviews Of Modern Physics}\ }\textbf {\bibinfo {volume} {82}},\ \bibinfo
  {pages} {3045} (\bibinfo {year} {2010})}\BibitemShut {NoStop}%
\bibitem [{\citenamefont {Khanikaev}\ \emph {et~al.}(2013)\citenamefont
  {Khanikaev}, \citenamefont {Mousavi},\ and\ \citenamefont
  {Tse}}]{Khanikaev2012}%
  \BibitemOpen
  \bibfield  {author} {\bibinfo {author} {\bibfnamefont {A.~B.}\ \bibnamefont
  {Khanikaev}}, \bibinfo {author} {\bibfnamefont {S.~H.}\ \bibnamefont
  {Mousavi}},\ and\ \bibinfo {author} {\bibfnamefont {W.-K.}\ \bibnamefont
  {Tse}},\ }\bibfield  {title} {\bibinfo {title} {Photonic topological
  insulators},\ }\href@noop {} {\bibfield  {journal} {\bibinfo  {journal}
  {Nature Materials}\ }\textbf {\bibinfo {volume} {12}},\ \bibinfo {pages}
  {233} (\bibinfo {year} {2013})}\BibitemShut {NoStop}%
\bibitem [{\citenamefont {Wu}\ and\ \citenamefont {Hu}(2015)}]{WuL2015}%
  \BibitemOpen
  \bibfield  {author} {\bibinfo {author} {\bibfnamefont {L.-H.}\ \bibnamefont
  {Wu}}\ and\ \bibinfo {author} {\bibfnamefont {X.}~\bibnamefont {Hu}},\
  }\bibfield  {title} {\bibinfo {title} {Scheme for achieving a topological
  photonic crystal by using dielectric material},\ }\href@noop {} {\bibfield
  {journal} {\bibinfo  {journal} {Physical Review Letters}\ }\textbf {\bibinfo
  {volume} {114}},\ \bibinfo {pages} {223901} (\bibinfo {year}
  {2015})}\BibitemShut {NoStop}%
\bibitem [{\citenamefont {Ozawa}\ \emph {et~al.}(2019)\citenamefont {Ozawa},
  \citenamefont {Price}, \citenamefont {Amo}, \citenamefont {Goldman},
  \citenamefont {Hafezi}, \citenamefont {Lu}, \citenamefont {Rechtsman},
  \citenamefont {Schuster}, \citenamefont {Simon}, \citenamefont {Zilberberg}
  \emph {et~al.}}]{Ozawa2019}%
  \BibitemOpen
  \bibfield  {author} {\bibinfo {author} {\bibfnamefont {T.}~\bibnamefont
  {Ozawa}}, \bibinfo {author} {\bibfnamefont {H.~M.}\ \bibnamefont {Price}},
  \bibinfo {author} {\bibfnamefont {A.}~\bibnamefont {Amo}}, \bibinfo {author}
  {\bibfnamefont {N.}~\bibnamefont {Goldman}}, \bibinfo {author} {\bibfnamefont
  {M.}~\bibnamefont {Hafezi}}, \bibinfo {author} {\bibfnamefont
  {L.}~\bibnamefont {Lu}}, \bibinfo {author} {\bibfnamefont {M.~C.}\
  \bibnamefont {Rechtsman}}, \bibinfo {author} {\bibfnamefont {D.}~\bibnamefont
  {Schuster}}, \bibinfo {author} {\bibfnamefont {J.}~\bibnamefont {Simon}},
  \bibinfo {author} {\bibfnamefont {O.}~\bibnamefont {Zilberberg}}, \emph
  {et~al.},\ }\bibfield  {title} {\bibinfo {title} {Topological photonics},\
  }\href@noop {} {\bibfield  {journal} {\bibinfo  {journal} {Reviews of Modern
  Physics}\ }\textbf {\bibinfo {volume} {91}},\ \bibinfo {pages} {015006}
  (\bibinfo {year} {2019})}\BibitemShut {NoStop}%
\bibitem [{\citenamefont {Price}\ \emph {et~al.}(2022)\citenamefont {Price},
  \citenamefont {Chong}, \citenamefont {Khanikaev}, \citenamefont {Schomerus},
  \citenamefont {Maczewsky}, \citenamefont {Kremer}, \citenamefont {Heinrich},
  \citenamefont {Szameit}, \citenamefont {Zilberberg}, \citenamefont {Yang},
  \citenamefont {Zhang}, \citenamefont {Alù}, \citenamefont {Thomale},
  \citenamefont {Carusotto}, \citenamefont {St-Jean}, \citenamefont {Amo},
  \citenamefont {Dutt}, \citenamefont {Yuan}, \citenamefont {Fan},
  \citenamefont {Yin}, \citenamefont {Peng}, \citenamefont {Ozawa},\ and\
  \citenamefont {Blanco-Redondo}}]{Price2022}%
  \BibitemOpen
  \bibfield  {author} {\bibinfo {author} {\bibfnamefont {H.}~\bibnamefont
  {Price}}, \bibinfo {author} {\bibfnamefont {Y.}~\bibnamefont {Chong}},
  \bibinfo {author} {\bibfnamefont {A.}~\bibnamefont {Khanikaev}}, \bibinfo
  {author} {\bibfnamefont {H.}~\bibnamefont {Schomerus}}, \bibinfo {author}
  {\bibfnamefont {L.~J.}\ \bibnamefont {Maczewsky}}, \bibinfo {author}
  {\bibfnamefont {M.}~\bibnamefont {Kremer}}, \bibinfo {author} {\bibfnamefont
  {M.}~\bibnamefont {Heinrich}}, \bibinfo {author} {\bibfnamefont
  {A.}~\bibnamefont {Szameit}}, \bibinfo {author} {\bibfnamefont
  {O.}~\bibnamefont {Zilberberg}}, \bibinfo {author} {\bibfnamefont
  {Y.}~\bibnamefont {Yang}}, \bibinfo {author} {\bibfnamefont {B.}~\bibnamefont
  {Zhang}}, \bibinfo {author} {\bibfnamefont {A.}~\bibnamefont {Alù}},
  \bibinfo {author} {\bibfnamefont {R.}~\bibnamefont {Thomale}}, \bibinfo
  {author} {\bibfnamefont {I.}~\bibnamefont {Carusotto}}, \bibinfo {author}
  {\bibfnamefont {P.}~\bibnamefont {St-Jean}}, \bibinfo {author} {\bibfnamefont
  {A.}~\bibnamefont {Amo}}, \bibinfo {author} {\bibfnamefont {A.}~\bibnamefont
  {Dutt}}, \bibinfo {author} {\bibfnamefont {L.}~\bibnamefont {Yuan}}, \bibinfo
  {author} {\bibfnamefont {S.}~\bibnamefont {Fan}}, \bibinfo {author}
  {\bibfnamefont {X.}~\bibnamefont {Yin}}, \bibinfo {author} {\bibfnamefont
  {C.}~\bibnamefont {Peng}}, \bibinfo {author} {\bibfnamefont {T.}~\bibnamefont
  {Ozawa}},\ and\ \bibinfo {author} {\bibfnamefont {A.}~\bibnamefont
  {Blanco-Redondo}},\ }\bibfield  {title} {\bibinfo {title} {Roadmap on
  topological photonics},\ }\href
  {http://iopscience.iop.org/article/10.1088/2515-7647/ac4ee4} {\bibfield
  {journal} {\bibinfo  {journal} {Journal of Physics: Photonics}\ } (\bibinfo
  {year} {2022})}\BibitemShut {NoStop}%
\bibitem [{\citenamefont {Wang}\ \emph {et~al.}(2009)\citenamefont {Wang},
  \citenamefont {Chong}, \citenamefont {Joannopoulos},\ and\ \citenamefont
  {Solja{\v c}i{\'c}}}]{WangZ2009}%
  \BibitemOpen
  \bibfield  {author} {\bibinfo {author} {\bibfnamefont {Z.}~\bibnamefont
  {Wang}}, \bibinfo {author} {\bibfnamefont {Y.}~\bibnamefont {Chong}},
  \bibinfo {author} {\bibfnamefont {J.~D.}\ \bibnamefont {Joannopoulos}},\ and\
  \bibinfo {author} {\bibfnamefont {M.}~\bibnamefont {Solja{\v c}i{\'c}}},\
  }\bibfield  {title} {\bibinfo {title} {Observation of unidirectional
  backscattering-immune topological electromagnetic states},\ }\href@noop {}
  {\bibfield  {journal} {\bibinfo  {journal} {Nature}\ }\textbf {\bibinfo
  {volume} {461}},\ \bibinfo {pages} {772} (\bibinfo {year}
  {2009})}\BibitemShut {NoStop}%
\bibitem [{\citenamefont {Fu}(2011)}]{Fu2011}%
  \BibitemOpen
  \bibfield  {author} {\bibinfo {author} {\bibfnamefont {L.}~\bibnamefont
  {Fu}},\ }\bibfield  {title} {\bibinfo {title} {Topological crystalline
  insulators},\ }\href@noop {} {\bibfield  {journal} {\bibinfo  {journal}
  {Physical Review Letters}\ }\textbf {\bibinfo {volume} {106}},\ \bibinfo
  {pages} {106802} (\bibinfo {year} {2011})}\BibitemShut {NoStop}%
\bibitem [{\citenamefont {Liang}\ and\ \citenamefont
  {Chong}(2013)}]{LiangG2013}%
  \BibitemOpen
  \bibfield  {author} {\bibinfo {author} {\bibfnamefont {G.}~\bibnamefont
  {Liang}}\ and\ \bibinfo {author} {\bibfnamefont {Y.}~\bibnamefont {Chong}},\
  }\bibfield  {title} {\bibinfo {title} {Optical resonator analog of a
  two-dimensional topological insulator},\ }\href@noop {} {\bibfield  {journal}
  {\bibinfo  {journal} {Physical Review Letters}\ }\textbf {\bibinfo {volume}
  {110}},\ \bibinfo {pages} {203904} (\bibinfo {year} {2013})}\BibitemShut
  {NoStop}%
\bibitem [{\citenamefont {Rechtsman}\ \emph {et~al.}(2013)\citenamefont
  {Rechtsman}, \citenamefont {Zeuner}, \citenamefont {Plotnik}, \citenamefont
  {Lumer}, \citenamefont {Podolsky}, \citenamefont {Dreisow}, \citenamefont
  {Nolte}, \citenamefont {Segev},\ and\ \citenamefont
  {Szameit}}]{Rechtsman2013}%
  \BibitemOpen
  \bibfield  {author} {\bibinfo {author} {\bibfnamefont {M.~C.}\ \bibnamefont
  {Rechtsman}}, \bibinfo {author} {\bibfnamefont {J.~M.}\ \bibnamefont
  {Zeuner}}, \bibinfo {author} {\bibfnamefont {Y.}~\bibnamefont {Plotnik}},
  \bibinfo {author} {\bibfnamefont {Y.}~\bibnamefont {Lumer}}, \bibinfo
  {author} {\bibfnamefont {D.}~\bibnamefont {Podolsky}}, \bibinfo {author}
  {\bibfnamefont {F.}~\bibnamefont {Dreisow}}, \bibinfo {author} {\bibfnamefont
  {S.}~\bibnamefont {Nolte}}, \bibinfo {author} {\bibfnamefont
  {M.}~\bibnamefont {Segev}},\ and\ \bibinfo {author} {\bibfnamefont
  {A.}~\bibnamefont {Szameit}},\ }\bibfield  {title} {\bibinfo {title}
  {Photonic floquet topological insulators},\ }\href
  {https://doi.org/10.1038/nature12066} {\bibfield  {journal} {\bibinfo
  {journal} {Nature}\ }\textbf {\bibinfo {volume} {496}},\ \bibinfo {pages}
  {196} (\bibinfo {year} {2013})}\BibitemShut {NoStop}%
\bibitem [{\citenamefont {Xu}\ \emph {et~al.}(2016)\citenamefont {Xu},
  \citenamefont {Wang}, \citenamefont {Xu}, \citenamefont {Chen},\ and\
  \citenamefont {Jiang}}]{XuL2016}%
  \BibitemOpen
  \bibfield  {author} {\bibinfo {author} {\bibfnamefont {L.}~\bibnamefont
  {Xu}}, \bibinfo {author} {\bibfnamefont {H.~X.}\ \bibnamefont {Wang}},
  \bibinfo {author} {\bibfnamefont {Y.~D.}\ \bibnamefont {Xu}}, \bibinfo
  {author} {\bibfnamefont {H.~Y.}\ \bibnamefont {Chen}},\ and\ \bibinfo
  {author} {\bibfnamefont {J.~H.}\ \bibnamefont {Jiang}},\ }\bibfield  {title}
  {\bibinfo {title} {Accidental degeneracy in photonic bands and topological
  phase transitions in two-dimensional core-shell dielectric photonic
  crystals},\ }\href {https://doi.org/10.1364/OE.24.018059} {\bibfield
  {journal} {\bibinfo  {journal} {Opt Express}\ }\textbf {\bibinfo {volume}
  {24}},\ \bibinfo {pages} {18059} (\bibinfo {year} {2016})}\BibitemShut
  {NoStop}%
\bibitem [{\citenamefont {Wen}\ \emph {et~al.}(2018)\citenamefont {Wen},
  \citenamefont {Qiu}, \citenamefont {Lu}, \citenamefont {He}, \citenamefont
  {Ke},\ and\ \citenamefont {Liu}}]{WenX2018}%
  \BibitemOpen
  \bibfield  {author} {\bibinfo {author} {\bibfnamefont {X.}~\bibnamefont
  {Wen}}, \bibinfo {author} {\bibfnamefont {C.}~\bibnamefont {Qiu}}, \bibinfo
  {author} {\bibfnamefont {J.}~\bibnamefont {Lu}}, \bibinfo {author}
  {\bibfnamefont {H.}~\bibnamefont {He}}, \bibinfo {author} {\bibfnamefont
  {M.}~\bibnamefont {Ke}},\ and\ \bibinfo {author} {\bibfnamefont
  {Z.}~\bibnamefont {Liu}},\ }\bibfield  {title} {\bibinfo {title} {Acoustic
  dirac degeneracy and topological phase transitions realized by rotating
  scatterers},\ }\href@noop {} {\bibfield  {journal} {\bibinfo  {journal}
  {Journal of Applied Physics}\ }\textbf {\bibinfo {volume} {123}},\ \bibinfo
  {pages} {091703} (\bibinfo {year} {2018})}\BibitemShut {NoStop}%
\bibitem [{\citenamefont {Geng}\ \emph {et~al.}(2019)\citenamefont {Geng},
  \citenamefont {Wang}, \citenamefont {Ma},\ and\ \citenamefont
  {Gao}}]{GengY2019}%
  \BibitemOpen
  \bibfield  {author} {\bibinfo {author} {\bibfnamefont {Y.~F.}\ \bibnamefont
  {Geng}}, \bibinfo {author} {\bibfnamefont {Z.~N.}\ \bibnamefont {Wang}},
  \bibinfo {author} {\bibfnamefont {Y.~G.}\ \bibnamefont {Ma}},\ and\ \bibinfo
  {author} {\bibfnamefont {F.}~\bibnamefont {Gao}},\ }\bibfield  {title}
  {\bibinfo {title} {Topological surface plasmon polaritons},\ }\href@noop {}
  {\bibfield  {journal} {\bibinfo  {journal} {Acta Physica Sinica}\ }\textbf
  {\bibinfo {volume} {68}},\ \bibinfo {pages} {224101} (\bibinfo {year}
  {2019})}\BibitemShut {NoStop}%
\bibitem [{\citenamefont {He}\ \emph {et~al.}(2020)\citenamefont {He},
  \citenamefont {Lai}, \citenamefont {He}, \citenamefont {Yu}, \citenamefont
  {Xu}, \citenamefont {Lu},\ and\ \citenamefont {Chen}}]{HeC2020}%
  \BibitemOpen
  \bibfield  {author} {\bibinfo {author} {\bibfnamefont {C.}~\bibnamefont
  {He}}, \bibinfo {author} {\bibfnamefont {H.~S.}\ \bibnamefont {Lai}},
  \bibinfo {author} {\bibfnamefont {B.}~\bibnamefont {He}}, \bibinfo {author}
  {\bibfnamefont {S.~Y.}\ \bibnamefont {Yu}}, \bibinfo {author} {\bibfnamefont
  {X.}~\bibnamefont {Xu}}, \bibinfo {author} {\bibfnamefont {M.~H.}\
  \bibnamefont {Lu}},\ and\ \bibinfo {author} {\bibfnamefont {Y.~F.}\
  \bibnamefont {Chen}},\ }\bibfield  {title} {\bibinfo {title} {Acoustic
  analogues of three-dimensional topological insulators},\ }\href
  {https://doi.org/10.1038/s41467-020-16131-w} {\bibfield  {journal} {\bibinfo
  {journal} {Nat Commun}\ }\textbf {\bibinfo {volume} {11}},\ \bibinfo {pages}
  {2318} (\bibinfo {year} {2020})}\BibitemShut {NoStop}%
\bibitem [{\citenamefont {Xie}\ \emph {et~al.}(2023)\citenamefont {Xie},
  \citenamefont {Jin},\ and\ \citenamefont {Song}}]{XieL2023}%
  \BibitemOpen
  \bibfield  {author} {\bibinfo {author} {\bibfnamefont {L.}~\bibnamefont
  {Xie}}, \bibinfo {author} {\bibfnamefont {L.}~\bibnamefont {Jin}},\ and\
  \bibinfo {author} {\bibfnamefont {Z.}~\bibnamefont {Song}},\ }\bibfield
  {title} {\bibinfo {title} {Antihelical edge states in two-dimensional
  photonic topological metals},\ }\href
  {https://doi.org/10.1016/j.scib.2023.01.018} {\bibfield  {journal} {\bibinfo
  {journal} {Sci Bull (Beijing)}\ }\textbf {\bibinfo {volume} {68}},\ \bibinfo
  {pages} {255} (\bibinfo {year} {2023})}\BibitemShut {NoStop}%
\bibitem [{\citenamefont {Zhou}\ \emph {et~al.}(2021)\citenamefont {Zhou},
  \citenamefont {Lin}, \citenamefont {Liu}, \citenamefont {Shi}, \citenamefont
  {Tang}, \citenamefont {Wu}, \citenamefont {Yu} \emph {et~al.}}]{ZhouR2021}%
  \BibitemOpen
  \bibfield  {author} {\bibinfo {author} {\bibfnamefont {R.}~\bibnamefont
  {Zhou}}, \bibinfo {author} {\bibfnamefont {H.}~\bibnamefont {Lin}}, \bibinfo
  {author} {\bibfnamefont {Y.}~\bibnamefont {Liu}}, \bibinfo {author}
  {\bibfnamefont {X.}~\bibnamefont {Shi}}, \bibinfo {author} {\bibfnamefont
  {R.}~\bibnamefont {Tang}}, \bibinfo {author} {\bibfnamefont {Y.}~\bibnamefont
  {Wu}}, \bibinfo {author} {\bibfnamefont {Z.}~\bibnamefont {Yu}}, \emph
  {et~al.},\ }\bibfield  {title} {\bibinfo {title} {Topological edge states of
  kekul{\'e}-type photonic crystals induced by a synchronized rotation of unit
  cells},\ }\href@noop {} {\bibfield  {journal} {\bibinfo  {journal} {Physical
  Review A}\ }\textbf {\bibinfo {volume} {104}},\ \bibinfo {pages} {L031502}
  (\bibinfo {year} {2021})}\BibitemShut {NoStop}%
\bibitem [{\citenamefont {Wang}\ and\ \citenamefont {Hu}(2020)}]{WangX2020}%
  \BibitemOpen
  \bibfield  {author} {\bibinfo {author} {\bibfnamefont {X.-X.}\ \bibnamefont
  {Wang}}\ and\ \bibinfo {author} {\bibfnamefont {X.}~\bibnamefont {Hu}},\
  }\bibfield  {title} {\bibinfo {title} {Reconfigurable topological waveguide
  based on honeycomb lattice of dielectric cuboids},\ }\href
  {https://doi.org/10.1515/nanoph-2020-0146} {\bibfield  {journal} {\bibinfo
  {journal} {Nanophotonics}\ }\textbf {\bibinfo {volume} {9}},\ \bibinfo
  {pages} {3451} (\bibinfo {year} {2020})}\BibitemShut {NoStop}%
\bibitem [{\citenamefont {Chen}\ \emph {et~al.}(2022)\citenamefont {Chen},
  \citenamefont {Li}, \citenamefont {Yu}, \citenamefont {Fu},\ and\
  \citenamefont {Hang}}]{ChenJ2022}%
  \BibitemOpen
  \bibfield  {author} {\bibinfo {author} {\bibfnamefont {J.-h.}\ \bibnamefont
  {Chen}}, \bibinfo {author} {\bibfnamefont {Y.}~\bibnamefont {Li}}, \bibinfo
  {author} {\bibfnamefont {C.}~\bibnamefont {Yu}}, \bibinfo {author}
  {\bibfnamefont {C.}~\bibnamefont {Fu}},\ and\ \bibinfo {author}
  {\bibfnamefont {Z.~H.}\ \bibnamefont {Hang}},\ }\bibfield  {title} {\bibinfo
  {title} {Realization of the quantum spin hall effect using tunable acoustic
  metamaterials},\ }\bibfield  {journal} {\bibinfo  {journal} {Physical Review
  Applied}\ }\textbf {\bibinfo {volume} {18}},\ \href
  {https://doi.org/10.1103/PhysRevApplied.18.044055}
  {10.1103/PhysRevApplied.18.044055} (\bibinfo {year} {2022})\BibitemShut
  {NoStop}%
\bibitem [{\citenamefont {Joannopoulos}\ \emph {et~al.}(2008)\citenamefont
  {Joannopoulos}, \citenamefont {Johnson}, \citenamefont {Winn},\ and\
  \citenamefont {Meade}}]{Joannopoulos2008}%
  \BibitemOpen
  \bibfield  {author} {\bibinfo {author} {\bibfnamefont {J.~D.}\ \bibnamefont
  {Joannopoulos}}, \bibinfo {author} {\bibfnamefont {S.~G.}\ \bibnamefont
  {Johnson}}, \bibinfo {author} {\bibfnamefont {J.~N.}\ \bibnamefont {Winn}},\
  and\ \bibinfo {author} {\bibfnamefont {R.~D.}\ \bibnamefont {Meade}},\
  }\href@noop {} {\emph {\bibinfo {title} {Photonic Crystals: Molding the Flow
  of Light}}},\ \bibinfo {edition} {second edition}\ ed.\ (\bibinfo
  {publisher} {Princeton University Press},\ \bibinfo {address} {Princeton and
  Oxford},\ \bibinfo {year} {2008})\BibitemShut {NoStop}%
\bibitem [{\citenamefont {Sui}\ \emph {et~al.}(2022)\citenamefont {Sui},
  \citenamefont {Liao}, \citenamefont {Li},\ and\ \citenamefont
  {Zhang}}]{SuiY2022}%
  \BibitemOpen
  \bibfield  {author} {\bibinfo {author} {\bibfnamefont {J.-Y.}\ \bibnamefont
  {Sui}}, \bibinfo {author} {\bibfnamefont {S.-y.}\ \bibnamefont {Liao}},
  \bibinfo {author} {\bibfnamefont {B.}~\bibnamefont {Li}},\ and\ \bibinfo
  {author} {\bibfnamefont {H.-F.}\ \bibnamefont {Zhang}},\ }\bibfield  {title}
  {\bibinfo {title} {High sensitivity multitasking non-reciprocity sensor using
  the photonic spin hall effect},\ }\href {https://doi.org/10.1364/OL.476048}
  {\bibfield  {journal} {\bibinfo  {journal} {Optics Letters}\ }\textbf
  {\bibinfo {volume} {47}},\ \bibinfo {pages} {6065} (\bibinfo {year}
  {2022})}\BibitemShut {NoStop}%
\bibitem [{\citenamefont {Zhou}\ \emph {et~al.}(2022)\citenamefont {Zhou},
  \citenamefont {Lin}, \citenamefont {Wu}, \citenamefont {Li}, \citenamefont
  {Yu}, \citenamefont {Liu},\ and\ \citenamefont {Xu}}]{ZhouR2022}%
  \BibitemOpen
  \bibfield  {author} {\bibinfo {author} {\bibfnamefont {R.}~\bibnamefont
  {Zhou}}, \bibinfo {author} {\bibfnamefont {H.}~\bibnamefont {Lin}}, \bibinfo
  {author} {\bibfnamefont {Y.}~\bibnamefont {Wu}}, \bibinfo {author}
  {\bibfnamefont {Z.}~\bibnamefont {Li}}, \bibinfo {author} {\bibfnamefont
  {Z.}~\bibnamefont {Yu}}, \bibinfo {author} {\bibfnamefont {Y.}~\bibnamefont
  {Liu}},\ and\ \bibinfo {author} {\bibfnamefont {D.-H.}\ \bibnamefont {Xu}},\
  }\bibfield  {title} {\bibinfo {title} {Higher-order valley vortices enabled
  by synchronized rotation in a photonic crystal},\ }\href@noop {} {\bibfield
  {journal} {\bibinfo  {journal} {Photonics Research}\ }\textbf {\bibinfo
  {volume} {10}},\ \bibinfo {pages} {1244} (\bibinfo {year}
  {2022})}\BibitemShut {NoStop}%
\bibitem [{\citenamefont {Yu}\ \emph {et~al.}(2022)\citenamefont {Yu},
  \citenamefont {Lin}, \citenamefont {Zhou}, \citenamefont {Li}, \citenamefont
  {Mao}, \citenamefont {Peng}, \citenamefont {Liu},\ and\ \citenamefont
  {Shi}}]{YuZ2022}%
  \BibitemOpen
  \bibfield  {author} {\bibinfo {author} {\bibfnamefont {Z.}~\bibnamefont
  {Yu}}, \bibinfo {author} {\bibfnamefont {H.}~\bibnamefont {Lin}}, \bibinfo
  {author} {\bibfnamefont {R.}~\bibnamefont {Zhou}}, \bibinfo {author}
  {\bibfnamefont {Z.}~\bibnamefont {Li}}, \bibinfo {author} {\bibfnamefont
  {Z.}~\bibnamefont {Mao}}, \bibinfo {author} {\bibfnamefont {K.}~\bibnamefont
  {Peng}}, \bibinfo {author} {\bibfnamefont {Y.}~\bibnamefont {Liu}},\ and\
  \bibinfo {author} {\bibfnamefont {X.}~\bibnamefont {Shi}},\ }\bibfield
  {title} {\bibinfo {title} {Topological valley crystals in a photonic
  {Su–Schrieffer–Heeger (SSH)} variant},\ }\bibfield  {journal} {\bibinfo
  {journal} {Journal of Applied Physics}\ }\textbf {\bibinfo {volume} {132}},\
  \href {https://doi.org/10.1063/5.0107211} {10.1063/5.0107211} (\bibinfo
  {year} {2022})\BibitemShut {NoStop}%
\bibitem [{\citenamefont {Yang}\ \emph {et~al.}(2018)\citenamefont {Yang},
  \citenamefont {Xu}, \citenamefont {Xu}, \citenamefont {Wang}, \citenamefont
  {Jiang}, \citenamefont {Hu},\ and\ \citenamefont {Hang}}]{YangY2018}%
  \BibitemOpen
  \bibfield  {author} {\bibinfo {author} {\bibfnamefont {Y.}~\bibnamefont
  {Yang}}, \bibinfo {author} {\bibfnamefont {Y.~F.}\ \bibnamefont {Xu}},
  \bibinfo {author} {\bibfnamefont {T.}~\bibnamefont {Xu}}, \bibinfo {author}
  {\bibfnamefont {H.-X.}\ \bibnamefont {Wang}}, \bibinfo {author}
  {\bibfnamefont {J.-H.}\ \bibnamefont {Jiang}}, \bibinfo {author}
  {\bibfnamefont {X.}~\bibnamefont {Hu}},\ and\ \bibinfo {author}
  {\bibfnamefont {Z.~H.}\ \bibnamefont {Hang}},\ }\bibfield  {title} {\bibinfo
  {title} {Visualization of a unidirectional electromagnetic waveguide using
  topological photonic crystals made of dielectric materials},\ }\href@noop {}
  {\bibfield  {journal} {\bibinfo  {journal} {Physical Review Letters}\
  }\textbf {\bibinfo {volume} {120}},\ \bibinfo {pages} {217401} (\bibinfo
  {year} {2018})}\BibitemShut {NoStop}%
\bibitem [{\citenamefont {Li}\ \emph {et~al.}(2009)\citenamefont {Li},
  \citenamefont {Chu}, \citenamefont {Jain},\ and\ \citenamefont
  {Shen}}]{LiJ2009}%
  \BibitemOpen
  \bibfield  {author} {\bibinfo {author} {\bibfnamefont {J.}~\bibnamefont
  {Li}}, \bibinfo {author} {\bibfnamefont {R.-L.}\ \bibnamefont {Chu}},
  \bibinfo {author} {\bibfnamefont {J.~K.}\ \bibnamefont {Jain}},\ and\
  \bibinfo {author} {\bibfnamefont {S.-Q.}\ \bibnamefont {Shen}},\ }\bibfield
  {title} {\bibinfo {title} {Topological {Anderson} insulator},\ }\href
  {https://doi.org/10.1103/PhysRevLett.102.136806} {\bibfield  {journal}
  {\bibinfo  {journal} {Physical Review Letters}\ }\textbf {\bibinfo {volume}
  {102}},\ \bibinfo {pages} {136806} (\bibinfo {year} {2009})}\BibitemShut
  {NoStop}%
\bibitem [{\citenamefont {Liu}\ \emph {et~al.}(2017)\citenamefont {Liu},
  \citenamefont {Gao}, \citenamefont {Yang},\ and\ \citenamefont
  {Zhang}}]{LiuC2017}%
  \BibitemOpen
  \bibfield  {author} {\bibinfo {author} {\bibfnamefont {C.}~\bibnamefont
  {Liu}}, \bibinfo {author} {\bibfnamefont {W.}~\bibnamefont {Gao}}, \bibinfo
  {author} {\bibfnamefont {B.}~\bibnamefont {Yang}},\ and\ \bibinfo {author}
  {\bibfnamefont {S.}~\bibnamefont {Zhang}},\ }\bibfield  {title} {\bibinfo
  {title} {Disorder-induced topological state transition in photonic
  metamaterials},\ }\href {https://doi.org/10.1103/PhysRevLett.119.183901}
  {\bibfield  {journal} {\bibinfo  {journal} {Phys Rev Lett}\ }\textbf
  {\bibinfo {volume} {119}},\ \bibinfo {pages} {183901} (\bibinfo {year}
  {2017})}\BibitemShut {NoStop}%
\bibitem [{\citenamefont {Stutzer}\ \emph {et~al.}(2018)\citenamefont
  {Stutzer}, \citenamefont {Plotnik}, \citenamefont {Lumer}, \citenamefont
  {Titum}, \citenamefont {Lindner}, \citenamefont {Segev}, \citenamefont
  {Rechtsman},\ and\ \citenamefont {Szameit}}]{Stutzer2018}%
  \BibitemOpen
  \bibfield  {author} {\bibinfo {author} {\bibfnamefont {S.}~\bibnamefont
  {Stutzer}}, \bibinfo {author} {\bibfnamefont {Y.}~\bibnamefont {Plotnik}},
  \bibinfo {author} {\bibfnamefont {Y.}~\bibnamefont {Lumer}}, \bibinfo
  {author} {\bibfnamefont {P.}~\bibnamefont {Titum}}, \bibinfo {author}
  {\bibfnamefont {N.~H.}\ \bibnamefont {Lindner}}, \bibinfo {author}
  {\bibfnamefont {M.}~\bibnamefont {Segev}}, \bibinfo {author} {\bibfnamefont
  {M.~C.}\ \bibnamefont {Rechtsman}},\ and\ \bibinfo {author} {\bibfnamefont
  {A.}~\bibnamefont {Szameit}},\ }\bibfield  {title} {\bibinfo {title}
  {Photonic topological anderson insulators},\ }\href
  {https://doi.org/10.1038/s41586-018-0418-2} {\bibfield  {journal} {\bibinfo
  {journal} {Nature}\ }\textbf {\bibinfo {volume} {560}},\ \bibinfo {pages}
  {461} (\bibinfo {year} {2018})}\BibitemShut {NoStop}%
\bibitem [{\citenamefont {Meier}\ \emph {et~al.}(2018)\citenamefont {Meier},
  \citenamefont {An}, \citenamefont {Dauphin}, \citenamefont {Maffei},
  \citenamefont {Massignan}, \citenamefont {Hughes},\ and\ \citenamefont
  {Gadway}}]{Meier2018}%
  \BibitemOpen
  \bibfield  {author} {\bibinfo {author} {\bibfnamefont {E.~J.}\ \bibnamefont
  {Meier}}, \bibinfo {author} {\bibfnamefont {F.~A.}\ \bibnamefont {An}},
  \bibinfo {author} {\bibfnamefont {A.}~\bibnamefont {Dauphin}}, \bibinfo
  {author} {\bibfnamefont {M.}~\bibnamefont {Maffei}}, \bibinfo {author}
  {\bibfnamefont {P.}~\bibnamefont {Massignan}}, \bibinfo {author}
  {\bibfnamefont {T.~L.}\ \bibnamefont {Hughes}},\ and\ \bibinfo {author}
  {\bibfnamefont {B.}~\bibnamefont {Gadway}},\ }\bibfield  {title} {\bibinfo
  {title} {Observation of the topological anderson insulator in disordered
  atomic wires},\ }\href {https://doi.org/10.1126/science.aat3406} {\bibfield
  {journal} {\bibinfo  {journal} {Science}\ }\textbf {\bibinfo {volume}
  {362}},\ \bibinfo {pages} {929} (\bibinfo {year} {2018})}\BibitemShut
  {NoStop}%
\bibitem [{\citenamefont {Liu}\ \emph {et~al.}(2020)\citenamefont {Liu},
  \citenamefont {Yang}, \citenamefont {Ren}, \citenamefont {Xue}, \citenamefont
  {Lin}, \citenamefont {Hu}, \citenamefont {Sun}, \citenamefont {Peng},
  \citenamefont {Zhou}, \citenamefont {Chong},\ and\ \citenamefont
  {Zhang}}]{LiuG2020}%
  \BibitemOpen
  \bibfield  {author} {\bibinfo {author} {\bibfnamefont {G.~G.}\ \bibnamefont
  {Liu}}, \bibinfo {author} {\bibfnamefont {Y.}~\bibnamefont {Yang}}, \bibinfo
  {author} {\bibfnamefont {X.}~\bibnamefont {Ren}}, \bibinfo {author}
  {\bibfnamefont {H.}~\bibnamefont {Xue}}, \bibinfo {author} {\bibfnamefont
  {X.}~\bibnamefont {Lin}}, \bibinfo {author} {\bibfnamefont {Y.~H.}\
  \bibnamefont {Hu}}, \bibinfo {author} {\bibfnamefont {H.~X.}\ \bibnamefont
  {Sun}}, \bibinfo {author} {\bibfnamefont {B.}~\bibnamefont {Peng}}, \bibinfo
  {author} {\bibfnamefont {P.}~\bibnamefont {Zhou}}, \bibinfo {author}
  {\bibfnamefont {Y.}~\bibnamefont {Chong}},\ and\ \bibinfo {author}
  {\bibfnamefont {B.}~\bibnamefont {Zhang}},\ }\bibfield  {title} {\bibinfo
  {title} {Topological anderson insulator in disordered photonic crystals},\
  }\href {https://doi.org/10.1103/PhysRevLett.125.133603} {\bibfield  {journal}
  {\bibinfo  {journal} {Phys Rev Lett}\ }\textbf {\bibinfo {volume} {125}},\
  \bibinfo {pages} {133603} (\bibinfo {year} {2020})}\BibitemShut {NoStop}%
\bibitem [{\citenamefont {Cui}\ \emph {et~al.}(2022)\citenamefont {Cui},
  \citenamefont {Zhang}, \citenamefont {Zhang},\ and\ \citenamefont
  {Chan}}]{CuiX2022}%
  \BibitemOpen
  \bibfield  {author} {\bibinfo {author} {\bibfnamefont {X.}~\bibnamefont
  {Cui}}, \bibinfo {author} {\bibfnamefont {R.~Y.}\ \bibnamefont {Zhang}},
  \bibinfo {author} {\bibfnamefont {Z.~Q.}\ \bibnamefont {Zhang}},\ and\
  \bibinfo {author} {\bibfnamefont {C.~T.}\ \bibnamefont {Chan}},\ }\bibfield
  {title} {\bibinfo {title} {Photonic $\mathbb{Z}_2$ topological anderson
  insulators},\ }\href {https://doi.org/10.1103/PhysRevLett.129.043902}
  {\bibfield  {journal} {\bibinfo  {journal} {Phys Rev Lett}\ }\textbf
  {\bibinfo {volume} {129}},\ \bibinfo {pages} {043902} (\bibinfo {year}
  {2022})}\BibitemShut {NoStop}%
\bibitem [{\citenamefont {Silveirinha}(2015)}]{Silveirinha2015}%
  \BibitemOpen
  \bibfield  {author} {\bibinfo {author} {\bibfnamefont {M.~G.}\ \bibnamefont
  {Silveirinha}},\ }\bibfield  {title} {\bibinfo {title} {Chern invariants for
  continuous media},\ }\bibfield  {journal} {\bibinfo  {journal} {Physical
  Review B}\ }\textbf {\bibinfo {volume} {92}},\ \href
  {https://doi.org/10.1103/PhysRevB.92.125153} {10.1103/PhysRevB.92.125153}
  (\bibinfo {year} {2015})\BibitemShut {NoStop}%
\bibitem [{\citenamefont {Chen}\ \emph {et~al.}(2017)\citenamefont {Chen},
  \citenamefont {Mei}, \citenamefont {Sun}, \citenamefont {Zhang},
  \citenamefont {Zhao},\ and\ \citenamefont {Wu}}]{ChenZ2017}%
  \BibitemOpen
  \bibfield  {author} {\bibinfo {author} {\bibfnamefont {Z.-G.}\ \bibnamefont
  {Chen}}, \bibinfo {author} {\bibfnamefont {J.}~\bibnamefont {Mei}}, \bibinfo
  {author} {\bibfnamefont {X.-C.}\ \bibnamefont {Sun}}, \bibinfo {author}
  {\bibfnamefont {X.}~\bibnamefont {Zhang}}, \bibinfo {author} {\bibfnamefont
  {J.}~\bibnamefont {Zhao}},\ and\ \bibinfo {author} {\bibfnamefont
  {Y.}~\bibnamefont {Wu}},\ }\bibfield  {title} {\bibinfo {title} {Multiple
  topological phase transitions in a gyromagnetic photonic crystal},\
  }\bibfield  {journal} {\bibinfo  {journal} {Physical Review A}\ }\textbf
  {\bibinfo {volume} {95}},\ \href {https://doi.org/10.1103/PhysRevA.95.043827}
  {10.1103/PhysRevA.95.043827} (\bibinfo {year} {2017})\BibitemShut {NoStop}%
\bibitem [{\citenamefont {Chen}\ \emph {et~al.}(2020)\citenamefont {Chen},
  \citenamefont {Liang},\ and\ \citenamefont {Li}}]{ChenJ2020}%
  \BibitemOpen
  \bibfield  {author} {\bibinfo {author} {\bibfnamefont {J.}~\bibnamefont
  {Chen}}, \bibinfo {author} {\bibfnamefont {W.}~\bibnamefont {Liang}},\ and\
  \bibinfo {author} {\bibfnamefont {Z.-Y.}\ \bibnamefont {Li}},\ }\bibfield
  {title} {\bibinfo {title} {Antichiral one-way edge states in a gyromagnetic
  photonic crystal},\ }\bibfield  {journal} {\bibinfo  {journal} {Physical
  Review B}\ }\textbf {\bibinfo {volume} {101}},\ \href
  {https://doi.org/10.1103/PhysRevB.101.214102} {10.1103/PhysRevB.101.214102}
  (\bibinfo {year} {2020})\BibitemShut {NoStop}%
\bibitem [{\citenamefont {Rao}\ \emph {et~al.}(2022)\citenamefont {Rao},
  \citenamefont {Fu}, \citenamefont {Zhang}, \citenamefont {Zhang},\ and\
  \citenamefont {Zhang}}]{RaoS2022}%
  \BibitemOpen
  \bibfield  {author} {\bibinfo {author} {\bibfnamefont {S.}~\bibnamefont
  {Rao}}, \bibinfo {author} {\bibfnamefont {H.}~\bibnamefont {Fu}}, \bibinfo
  {author} {\bibfnamefont {J.}~\bibnamefont {Zhang}}, \bibinfo {author}
  {\bibfnamefont {D.}~\bibnamefont {Zhang}},\ and\ \bibinfo {author}
  {\bibfnamefont {H.}~\bibnamefont {Zhang}},\ }\bibfield  {title} {\bibinfo
  {title} {Tunable phase retarder of 1d layered photonic structure combining
  nonlinear kerr dielectric defect layers and magnetized plasma materials},\
  }\href {https://doi.org/https://doi.org/10.1002/andp.202200337} {\bibfield
  {journal} {\bibinfo  {journal} {Annalen der Physik}\ }\textbf {\bibinfo
  {volume} {534}},\ \bibinfo {pages} {2200337} (\bibinfo {year}
  {2022})}\BibitemShut {NoStop}%
\bibitem [{\citenamefont {Sui}\ \emph {et~al.}(2023)\citenamefont {Sui},
  \citenamefont {Dong}, \citenamefont {Liao}, \citenamefont {Zhao},
  \citenamefont {Wang},\ and\ \citenamefont {Zhang}}]{SuiY2023}%
  \BibitemOpen
  \bibfield  {author} {\bibinfo {author} {\bibfnamefont {J.}~\bibnamefont
  {Sui}}, \bibinfo {author} {\bibfnamefont {R.}~\bibnamefont {Dong}}, \bibinfo
  {author} {\bibfnamefont {S.}~\bibnamefont {Liao}}, \bibinfo {author}
  {\bibfnamefont {Z.}~\bibnamefont {Zhao}}, \bibinfo {author} {\bibfnamefont
  {Y.}~\bibnamefont {Wang}},\ and\ \bibinfo {author} {\bibfnamefont {H.-F.}\
  \bibnamefont {Zhang}},\ }\bibfield  {title} {\bibinfo {title} {Janus
  metastructure based on magnetized plasma material with and logic gate and
  multiple physical quantity detection},\ }\href
  {https://doi.org/https://doi.org/10.1002/andp.202200509} {\bibfield
  {journal} {\bibinfo  {journal} {Annalen der Physik}\ }\textbf {\bibinfo
  {volume} {535}},\ \bibinfo {pages} {2200509} (\bibinfo {year}
  {2023})}\BibitemShut {NoStop}%
\bibitem [{\citenamefont {Zhao}\ \emph {et~al.}(2020)\citenamefont {Zhao},
  \citenamefont {Xie}, \citenamefont {Chen}, \citenamefont {Lan}, \citenamefont
  {Huang},\ and\ \citenamefont {Sha}}]{ZhaoR2020}%
  \BibitemOpen
  \bibfield  {author} {\bibinfo {author} {\bibfnamefont {R.}~\bibnamefont
  {Zhao}}, \bibinfo {author} {\bibfnamefont {G.~D.}\ \bibnamefont {Xie}},
  \bibinfo {author} {\bibfnamefont {M.~L.~N.}\ \bibnamefont {Chen}}, \bibinfo
  {author} {\bibfnamefont {Z.}~\bibnamefont {Lan}}, \bibinfo {author}
  {\bibfnamefont {Z.}~\bibnamefont {Huang}},\ and\ \bibinfo {author}
  {\bibfnamefont {W.~E.~I.}\ \bibnamefont {Sha}},\ }\bibfield  {title}
  {\bibinfo {title} {First-principle calculation of chern number in gyrotropic
  photonic crystals},\ }\href {https://doi.org/10.1364/OE.380077} {\bibfield
  {journal} {\bibinfo  {journal} {Opt Express}\ }\textbf {\bibinfo {volume}
  {28}},\ \bibinfo {pages} {4638} (\bibinfo {year} {2020})}\BibitemShut
  {NoStop}%
\bibitem [{\citenamefont {Wang}\ \emph {et~al.}(2019)\citenamefont {Wang},
  \citenamefont {Guo},\ and\ \citenamefont {Jiang}}]{Haixiao2020}%
  \BibitemOpen
  \bibfield  {author} {\bibinfo {author} {\bibfnamefont {H.-X.}\ \bibnamefont
  {Wang}}, \bibinfo {author} {\bibfnamefont {G.-Y.}\ \bibnamefont {Guo}},\ and\
  \bibinfo {author} {\bibfnamefont {J.-H.}\ \bibnamefont {Jiang}},\ }\bibfield
  {title} {\bibinfo {title} {Band topology in classical waves: Wilson-loop
  approach to topological numbers and fragile topology},\ }\bibfield  {journal}
  {\bibinfo  {journal} {New Journal of Physics}\ }\textbf {\bibinfo {volume}
  {21}},\ \href {https://doi.org/10.1088/1367-2630/ab3f71}
  {10.1088/1367-2630/ab3f71} (\bibinfo {year} {2019})\BibitemShut {NoStop}%
\bibitem [{\citenamefont {Fukui}\ \emph {et~al.}(2005)\citenamefont {Fukui},
  \citenamefont {Hatsugai},\ and\ \citenamefont {Suzuki}}]{Fukui2005}%
  \BibitemOpen
  \bibfield  {author} {\bibinfo {author} {\bibfnamefont {T.}~\bibnamefont
  {Fukui}}, \bibinfo {author} {\bibfnamefont {Y.}~\bibnamefont {Hatsugai}},\
  and\ \bibinfo {author} {\bibfnamefont {H.}~\bibnamefont {Suzuki}},\
  }\bibfield  {title} {\bibinfo {title} {Chern numbers in discretized brillouin
  zone: Efficient method of computing (spin) hall conductances},\ }\href
  {https://doi.org/10.1143/jpsj.74.1674} {\bibfield  {journal} {\bibinfo
  {journal} {Journal of the Physical Society of Japan}\ }\textbf {\bibinfo
  {volume} {74}},\ \bibinfo {pages} {1674} (\bibinfo {year}
  {2005})}\BibitemShut {NoStop}%
\bibitem [{\citenamefont {Jin}\ \emph {et~al.}(2017)\citenamefont {Jin},
  \citenamefont {Christensen}, \citenamefont {Soljacic}, \citenamefont {Fang},
  \citenamefont {Lu},\ and\ \citenamefont {Zhang}}]{JinD2017}%
  \BibitemOpen
  \bibfield  {author} {\bibinfo {author} {\bibfnamefont {D.}~\bibnamefont
  {Jin}}, \bibinfo {author} {\bibfnamefont {T.}~\bibnamefont {Christensen}},
  \bibinfo {author} {\bibfnamefont {M.}~\bibnamefont {Soljacic}}, \bibinfo
  {author} {\bibfnamefont {N.~X.}\ \bibnamefont {Fang}}, \bibinfo {author}
  {\bibfnamefont {L.}~\bibnamefont {Lu}},\ and\ \bibinfo {author}
  {\bibfnamefont {X.}~\bibnamefont {Zhang}},\ }\bibfield  {title} {\bibinfo
  {title} {Infrared topological plasmons in graphene},\ }\href
  {https://doi.org/10.1103/PhysRevLett.118.245301} {\bibfield  {journal}
  {\bibinfo  {journal} {Phys Rev Lett}\ }\textbf {\bibinfo {volume} {118}},\
  \bibinfo {pages} {245301} (\bibinfo {year} {2017})}\BibitemShut {NoStop}%
\bibitem [{\citenamefont {Skirlo}\ \emph {et~al.}(2014)\citenamefont {Skirlo},
  \citenamefont {Lu},\ and\ \citenamefont {Soljacic}}]{Skirlo2014}%
  \BibitemOpen
  \bibfield  {author} {\bibinfo {author} {\bibfnamefont {S.~A.}\ \bibnamefont
  {Skirlo}}, \bibinfo {author} {\bibfnamefont {L.}~\bibnamefont {Lu}},\ and\
  \bibinfo {author} {\bibfnamefont {M.}~\bibnamefont {Soljacic}},\ }\bibfield
  {title} {\bibinfo {title} {Multimode one-way waveguides of large chern
  numbers},\ }\href {https://doi.org/10.1103/PhysRevLett.113.113904} {\bibfield
   {journal} {\bibinfo  {journal} {Phys Rev Lett}\ }\textbf {\bibinfo {volume}
  {113}},\ \bibinfo {pages} {113904} (\bibinfo {year} {2014})}\BibitemShut
  {NoStop}%
\bibitem [{\citenamefont {Skirlo}\ \emph {et~al.}(2015)\citenamefont {Skirlo},
  \citenamefont {Lu}, \citenamefont {Igarashi}, \citenamefont {Yan},
  \citenamefont {Joannopoulos},\ and\ \citenamefont {Soljacic}}]{Skirlo2015}%
  \BibitemOpen
  \bibfield  {author} {\bibinfo {author} {\bibfnamefont {S.~A.}\ \bibnamefont
  {Skirlo}}, \bibinfo {author} {\bibfnamefont {L.}~\bibnamefont {Lu}}, \bibinfo
  {author} {\bibfnamefont {Y.}~\bibnamefont {Igarashi}}, \bibinfo {author}
  {\bibfnamefont {Q.}~\bibnamefont {Yan}}, \bibinfo {author} {\bibfnamefont
  {J.}~\bibnamefont {Joannopoulos}},\ and\ \bibinfo {author} {\bibfnamefont
  {M.}~\bibnamefont {Soljacic}},\ }\bibfield  {title} {\bibinfo {title}
  {Experimental observation of large chern numbers in photonic crystals},\
  }\href {https://doi.org/10.1103/PhysRevLett.115.253901} {\bibfield  {journal}
  {\bibinfo  {journal} {Phys Rev Lett}\ }\textbf {\bibinfo {volume} {115}},\
  \bibinfo {pages} {253901} (\bibinfo {year} {2015})}\BibitemShut {NoStop}%
\bibitem [{\citenamefont {Tian}(2023)}]{TianThesis2023}%
  \BibitemOpen
  \bibfield  {author} {\bibinfo {author} {\bibfnamefont {Y.}~\bibnamefont
  {Tian}},\ }\emph {\bibinfo {title} {Breakdown of topological phase due to
  unit-cell disorder in a gyromagnetic photonic crystal}},\ \href@noop {}
  {\bibinfo {type} {Master thesis}} (\bibinfo {year} {2023})\BibitemShut
  {NoStop}%
\bibitem [{\citenamefont {Zhang}\ \emph {et~al.}(2022)\citenamefont {Zhang},
  \citenamefont {Sui}, \citenamefont {Zhang}, \citenamefont {Liu},
  \citenamefont {Shi}, \citenamefont {Lv}, \citenamefont {Zhang}, \citenamefont
  {Rong},\ and\ \citenamefont {Yang}}]{ZhangH2022}%
  \BibitemOpen
  \bibfield  {author} {\bibinfo {author} {\bibfnamefont {H.}~\bibnamefont
  {Zhang}}, \bibinfo {author} {\bibfnamefont {W.}~\bibnamefont {Sui}}, \bibinfo
  {author} {\bibfnamefont {Y.}~\bibnamefont {Zhang}}, \bibinfo {author}
  {\bibfnamefont {G.}~\bibnamefont {Liu}}, \bibinfo {author} {\bibfnamefont
  {Q.}~\bibnamefont {Shi}}, \bibinfo {author} {\bibfnamefont {Z.}~\bibnamefont
  {Lv}}, \bibinfo {author} {\bibfnamefont {D.}~\bibnamefont {Zhang}}, \bibinfo
  {author} {\bibfnamefont {C.}~\bibnamefont {Rong}},\ and\ \bibinfo {author}
  {\bibfnamefont {B.}~\bibnamefont {Yang}},\ }\bibfield  {title} {\bibinfo
  {title} {Disorder‐driven collapse of topological phases in photonic
  topological insulator},\ }\bibfield  {journal} {\bibinfo  {journal} {Physica
  Status Solidi B}\ }\textbf {\bibinfo {volume} {259}},\ \href
  {https://doi.org/10.1002/pssb.202200214} {10.1002/pssb.202200214} (\bibinfo
  {year} {2022})\BibitemShut {NoStop}%
\end{thebibliography}%



\end{document}